\documentclass[aps,pre,superscriptaddress,twocolumn]{revtex4}
\usepackage[dvips]{graphics}
\usepackage{graphicx}
\usepackage{amsfonts}
\usepackage{amssymb}
\usepackage{amsmath}
\usepackage{subfigure}

\begin{document}

\title{Dark Solitary Waves in a Class of Collisionally Inhomogeneous Bose-Einstein Condensates}

\author{Chang Wang}
\affiliation{Oxford Centre for Industrial and Applied Mathematics, Mathematical Institute, University of Oxford, OX1 3LB, UK}

\author{Kody J. H. Law}
\affiliation{Mathematics Institute, University of Warwick, CV4 7AL, UK}

\author{Panayotis. G. Kevrekidis }
\affiliation{Department of Mathematics and Statistics, University of Massachusetts,
Amherst, MA 01003-4515, USA}

\author{Mason A. Porter}
\affiliation{Oxford Centre for Industrial and Applied Mathematics, Mathematical Institute, University of Oxford, OX1 3LB, UK}



\begin{abstract}

We study the structure, stability, and dynamics of dark solitary waves in parabolically trapped, collisionally inhomogeneous Bose-Einstein condensates with spatially periodic variations of the scattering length.  This collisional inhomogeneity yields a nonlinear lattice, which we tune from a small-amplitude, approximately sinusoidal structure to a periodic sequence of densely-spaced spikes.  We start by investigating time-independent inhomogeneities, and we subsequently examine the dynamical response when one starts with a collisionally homogeneous BEC and 
then switches on an inhomogeneity either adiabatically or nonadiabatically.  Using Bogoliubov-de Gennes linearization as well as direct numerical simulations of the Gross-Pitaevskii equation, we observe dark solitary waves, which can become unstable through oscillatory or exponential instabilities.  We find a critical wavelength of the nonlinear lattice that is comparable
to the healing length. Near this value, the fundamental eigenmode responsible for the stability of the dark solitary wave changes its direction of movement as a function of the strength of the nonlinearity. When it increases, it collides with other eigenmodes, leading to oscillatory instabilities; when it decreases, it collides with the origin and becomes imaginary, illustrating that the instability
mechanism is fundamentally different in wide-well versus narrow-well lattices. When starting from a collisionally homogeneous setup and switching on inhomogeneities, we find that dark solitary waves are preserved generically for aligned lattices.  We briefly examine the time scales for the onset of solitary-wave oscillations in this scenario.

\end{abstract}

\maketitle


\section{Introduction}

Because of advances in the field of Bose-Einstein condensation, solitary waves have received considerable attention in atomic physics during the past 15 years \cite{book1,book2}.  In particular, many studies have examined a mean-field description of atomic Bose-Einstein condensates (BECs) using the Gross-Pitaevskii (GP) equation. The GP equation includes a cubic nonlinearity, with a strength proportional to the $s$-wave scattering length, which arises from the interatomic interactions. This mean-field description has been important for theoretical and experiment investigations of phenomena such as bright solitary waves \cite{expb1,expb2,expb3}, dark solitary waves \cite{dark1,dark2,dark3,dark4,dark5,dark6,dark7,dark8}, gap solitary waves \cite{gap}, multi-component solitary waves \cite{db1,db2,db3,db4,hoefer,ourbook,review}, and Faraday waves \cite{engels07,alex1,alex2}.  Solitary waves have been studied in great detail in the presence of various external potentials, including linear periodic lattices (so-called ``optical lattices") and harmonic traps. This has yielded tremendous insights into a large 
variety of phenomena, including Bloch oscillations, Landau-Zener tunneling, modulational (``dynamical") instabilities and gap excitations, and more \cite{ourbook,review,pgk,konotop,blochol,morsch}. 

Numerous techniques have been developed that enable the control and manipulation of coherent states in BECs.  These include static (homogeneous and inhomogeneous) electric and magnetic 
fields \cite{Folman}, optical devices \cite{Grimm}, and near-field 
radio-frequency devices \cite{Lesanovsky}.  One can vary a BEC's 
external (trapping) potential while independently and simultaneously changing the strength of the 
nonlinearity by tuning interatomic interactions.  The interaction among the atoms can 
be adjusted experimentally over a very broad range by employing either 
magnetic \cite{Koehler,feshbachNa} or optical Feshbach resonances \cite{ofr}, and the manipulation of BECs using Feshbach resonances has led to numerous insights.  Experimental achievements include the formation of bright solitary waves and solitary-wave trains for $^{7}$Li \cite{expb1,expb2} and $^{85}$Rb \cite{expb3} atoms by tuning the interatomic interaction within a stable BEC from repulsive to attractive, the formation of molecular condensates \cite{molecule}, and the probing of the BEC-BCS crossover \cite{becbcs}.  Theoretical studies include the prediction that a time-dependent modulation of the scattering length can be used to stabilize attractive two-dimensional (2D) BECs against collapse \cite{FRM1} or to create robust matter-wave breathers in 1D BECs \cite{FRM2}. Temporal modulation of the GP equation's nonlinearity and its impact
on collapse properties and on other nonlinear phenomena (such as modulational instabilities) have been examined systematically in nonlinear optics~\cite{martin}.

Atomic matter waves also exhibit novel features under the influence of a spatially varying scattering length, which yields a spatially-dependent nonlinearity coefficient in the GP equation.  Numerous works have considered matter-wave dynamics in such ``collisionally inhomogeneous'' 
environments.  Theoretical predictions include adiabatic compression of matter waves \cite{our1,fka}, enhancement of the transmitivity of matter waves through barriers \cite{our2,fka2}, dynamical trapping of solitary waves \cite{our2}, and a delocalization transition of matter waves \cite{LocDeloc}.  Linear \cite{our1,our2}, parabolic \cite{yiota}, random \cite{vpg14}, periodic (i.e., nonlinear lattices) \cite{vpg16,LocDeloc,BludKon,rodrig08}, and localized (step-like) \cite{vpg12,vpg17,vpg_new} inhomogeneities have all been considered.  There have also been several detailed mathematical studies \cite{key-2,key-4,vprl} as well as examinations of analogous situations in optics \cite{kominis}.  Additionally, the interplay between linear and nonlinear lattices has been examined in both continuum \cite{ckrtj} and discrete \cite{blud_pre} settings. Two important recent developments on BECs in collisionally inhomogeneous environments include a broad review of relevant theoretical activity \cite{malomed} and the experimental creation of a spatially periodic effective nonlinearity in a Yb BEC \cite{takahashi}.

In this paper, which is motivated in part by the recent experimental
work of Ref.~\cite{takahashi}, we consider spatially periodic scattering lengths 
that can be tuned using Feshbach resonances from small-amplitude, 
approximately sinusoidal structures to a periodic sequence of densely-spaced 
spikes.  We consider ``aligned" and ``anti-aligned" periodic 
structures and compare and contrast the solitary-wave dynamics for each case.  
We first consider situations in which collisional inhomogeneities are always 
present (i.e., in which the coefficient of the nonlinearity is time-independent) 
before moving on to ones in which they are turned on either 
adiabatically or abruptly.  

When considering time-independent inhomogeneities, we employ both direct numerical simulations of the GP equation and a Bogoliubov-de Gennes (BdG) analysis to examine the linear stability of dark solitary waves.  We thereby identify two distinct mechanisms of instability for the dark solitary waves---an exponential one and an oscillatory one---and we examine when each of
these arises. Importantly, we find and quantify a change of instability mechanism as one switches between wide-well and narrow-well lattices. When switching on inhomogeneities in time, we find for aligned lattices that the dark solitary waves are generically robust, but that additional excitations in the form of gray solitary waves also emerge (especially when the inhomogeneities are turned on 
nonadiabatically).  We briefly examine the time scales for the onset of solitary-wave oscillations in this scenario.

The rest of our presentation is organized as follows.  In Section \ref{setup}, we present the GP equation and our setup of spatially periodic scattering lengths.  We consider time-independent scattering lengths and discuss the dark solitary waves arising in this situation in Section \ref{dark}.  We then consider the dynamical response of switching on the collisional inhomogeneities in Section \ref{response}.  Finally, we conclude in Section \ref{conclusion} and present some directions for future work.


\section{Setup and Model} \label{setup}

The dynamics of a cigar-shaped BEC can be approximated in 
the mean field by the quasi-1D GP equation \cite{book1,book2,ourbook}
\begin{align}
	i \frac{\partial}{\partial t} u (z,t) = \left( -\frac{\partial^2}{\partial z^2} + V(z,t) + 
g(z,t) |u(z,t)| ^{2} - \mu \right) u(z,t)\,, \label{eq:basicGP}
\end{align}
where $u(z,t)$ is the macroscopic wave function, $\mu$ is the chemical potential \footnote{Note that $\tilde{u}(z,t) = e^{-i\mu t}u(z,t)$ satisfies the quasi-1D GP equation without the need to write the chemical potential $\mu$ explicitly in the equation.},
$V(z,t)$ is an external potential, and $g(z,t)$ is a (suitably normalized) spatially and temporally modulated coefficient. We suppose that the BEC is in a harmonic trap, so 
$V(z,t) = \frac{1}{4 B^2} z^2$. In Eq.~(\ref{eq:basicGP}), we have rescaled the condensate density in units of $B \hbar \omega_z / |q|$, length in
units of $a_{z}/\sqrt{2B}$, time in units of $(B \omega_{z})^{-1}$, and energy in units of $\hbar B \omega_{z}$. The parameter $a_{z} = \sqrt{\hbar/(m \omega_z)}$ is the axial
oscillator length, $\hbar$ is Planck's constant, $m$ is the mass of an atom in the BEC, and $\omega_{z}$ is the axial trap frequency.  Additionally, $q=2 a \hbar \omega_\perp$ is the 
dimensional nonlinearity coefficient, where  $a$ is the $s$-wave
scattering length and we have averaged over the transverse directions 
(assuming a ground-state wave profile) \cite{ourbook,review}. We choose $B = 10$ for convenience. We consider a $^{87}$Rb BEC as an example and consequently use the parameter values $a \approx 5.5 \times 10^{-9}$m, $\omega_z=2\pi \times 4 s^{-1}$, and $\omega_\perp=80 \omega_z$.

In our calculations, the nondimensional numbers of atoms $N = \int_{-\infty}^{+\infty} |u|^2 dz$ is roughly 
$500$.  The number of atoms $\mathcal{N} = ({\hbar}^{3/2} [ B \omega_z / 2 m ]^{1/2} / |q|) N$ is thus roughly 
$7000$, though the phenomena that we observe are robust over different values of $N$. 
The two criteria for the quasi-1D regime are
satisfied \cite{ds_dimitri}: (1) the trap is highly anisotropic, as
$\Omega \equiv \omega_z / \omega_{\perp} = 0.0125 \ll 1$; and (2) the 
parameter $d \equiv \mathcal{N} \Omega a/a_\perp \approx 0.8 < 1$, where $a_{\perp} = \sqrt{\hbar/(m \omega_\perp)}$ is the oscillator length in the radial direction.

The normalized coefficient $g \propto a$ can be either positive or negative.  
The sign of $g$ depends on the atomic species; repulsive interatomic interactions yield $g > 0$, and attractive ones yield $g < 0$.  The sign and magnitude of $g$ can both be changed using 
Feshbach resonances, which make it possible (in principle) to manipulate the sign and strength of atomic interactions \cite{Koehler,feshbachNa}.  It is also possible to vary 
the scattering length and consequently the nonlinearity coefficient
$g$ in space and time by tuning an external field in the vicinity of a 
Feshbach resonance.  As we noted previously, spatial variation of the $s$-wave scattering length using Feshbach resonances was recently demonstrated experimentally \cite{takahashi}.  Temporal variations of BEC scattering lengths using Feshbach resonances have been studied for more than 10 years. 

In this paper, we consider repulsive BECs with a spatially periodic nonlinearity coefficient $g = g(z,t)$.  We assume that the nonlinearity does not change sign. We start by considering time-independent $s$-wave scattering lengths, for which $g(z,t) = g(z)$.  We consider the experimentally realistic \cite{takahashi} functional form
\begin{align}
    	g(z) & = g_0 + \Delta g(z)\,, \notag \\
 	\Delta g(z) &= \frac{g_m}{1+g_s \sin(kz+\phi)}\,,  \label{modeldeltag}
\end{align}
where the wavenumber $k$ determines the wavelength of the nonlinear lattice $g(z)$, the parameter $g_0=1$ (by normalization), and $\Delta g(z) \geq 0$.  We let $g_m \in [0,1]$, and we use $g_s \in [0,0.95]$ so that $\Delta g(z)$ is always finite.  In our numerical simulations, we consider wavenumbers $k \in [0,15]$.

The parameter $g_m$ determines the magnitude of the spatial modulation
of $g(z)$, and the parameter $g_s$ determines the shape of the lattice, which is
approximately sinusoidal when $g_s \ll 1$ and resembles a periodic
train of thin spikes as $g_s \rightarrow 1$.  We avoid a singularity in $g(z)$ by bounding $g_s$ away from $1$. The angle $\phi$ is a constant and determines a shift in alignment between the nonlinear lattice and the harmonic potential $V(z)$. We consider two situations: (1) \emph{aligned} nonlinear lattices, for which $\phi = \frac{\pi}{2}$, so the minimum of the harmonic potential $V(z)$ (which is located at $z = 0$) coincides with a minimum of $g(z)$; and (2) \emph{anti-aligned} lattices, for which $\phi=-\frac{\pi}{2}$, so the minimum of the harmonic trap coincides with a maximum of $g(z)$.  As we discuss later, the stability properties of solitary waves differ in the two cases.

In the sections below, we examine the existence, stability, and dynamics of dark solitary waves as the strength and shape of the nonlinear lattice $g(z)$ is varied. We establish existence by finding standing-wave solutions $\bar{u}(z)$, for which the right-hand side of Eq.~(\ref{eq:basicGP}) vanishes. We then examine the stability of these solutions via linearization around $\bar{u}$. In other words, we perform a BdG analysis: we consider an $O(\varepsilon)$ correction to the GP equation (\ref{eq:basicGP}) by writing
\begin{equation*}
	u(z,t)= \bar{u}(z) + \varepsilon\left[ a(z) e^{-i \omega t}+b^{*}(z) e^{i \omega t}\right]\,, 
\end{equation*}	
where the asterisk $^{*}$ denotes complex conjugation. We examine the dynamics of the standing wave by perturbing it and computing its temporal evolution using Eq.~(\ref{eq:basicGP}). 
We then study the response of solitary waves to both gradual (adiabatic) and 
abrupt (nonadiabatic) changes of the atomic interactions.  Both types of 
changes can, in principle, be introduced experimentally using a Feshbach resonance.


\section{Dark Solitary Waves} \label{dark}

\begin{figure}[ht]
\centering
\includegraphics[width=0.22\textwidth]{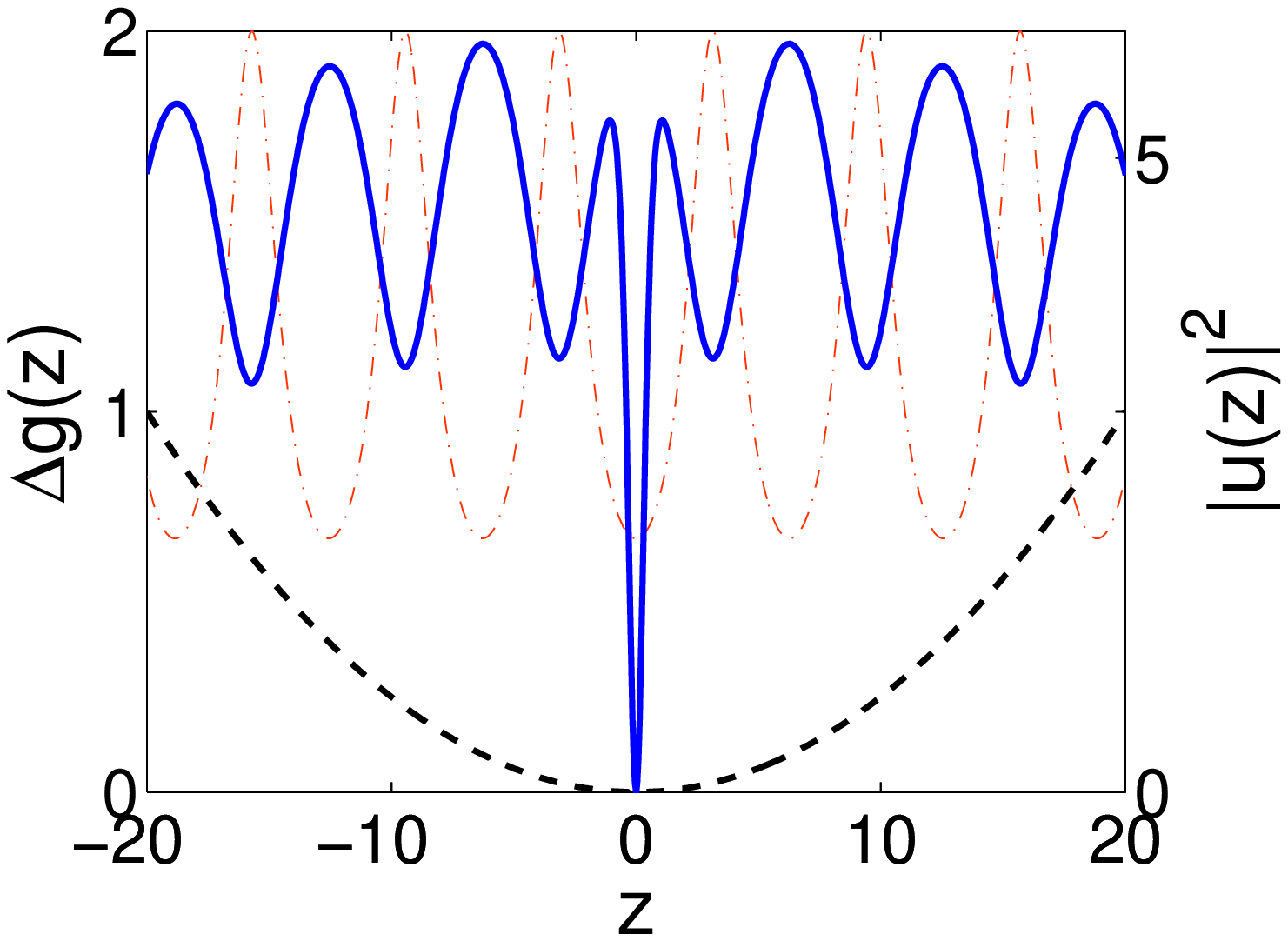}
\includegraphics[width=0.22\textwidth]{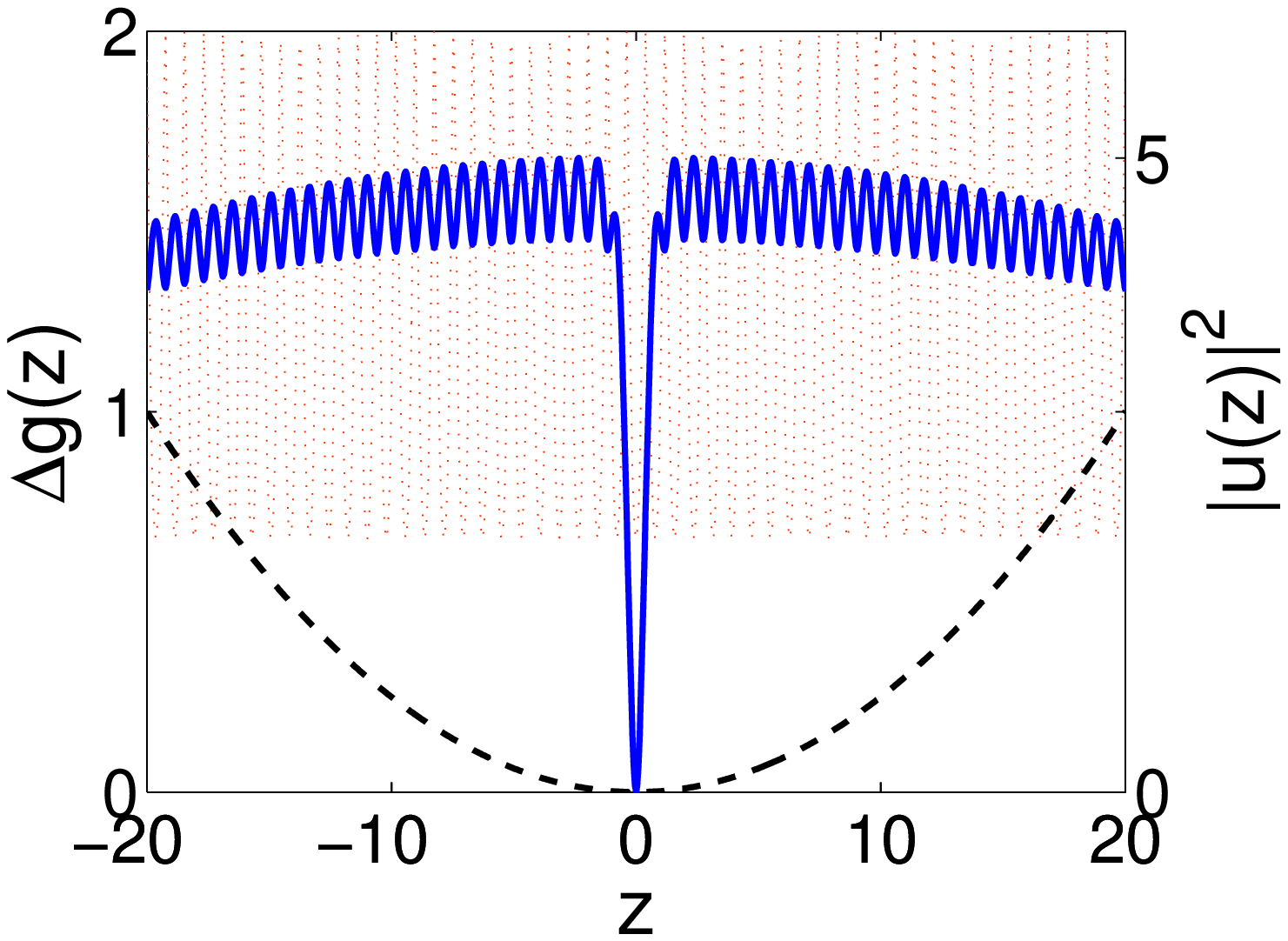}\\
\includegraphics[width=0.22\textwidth]{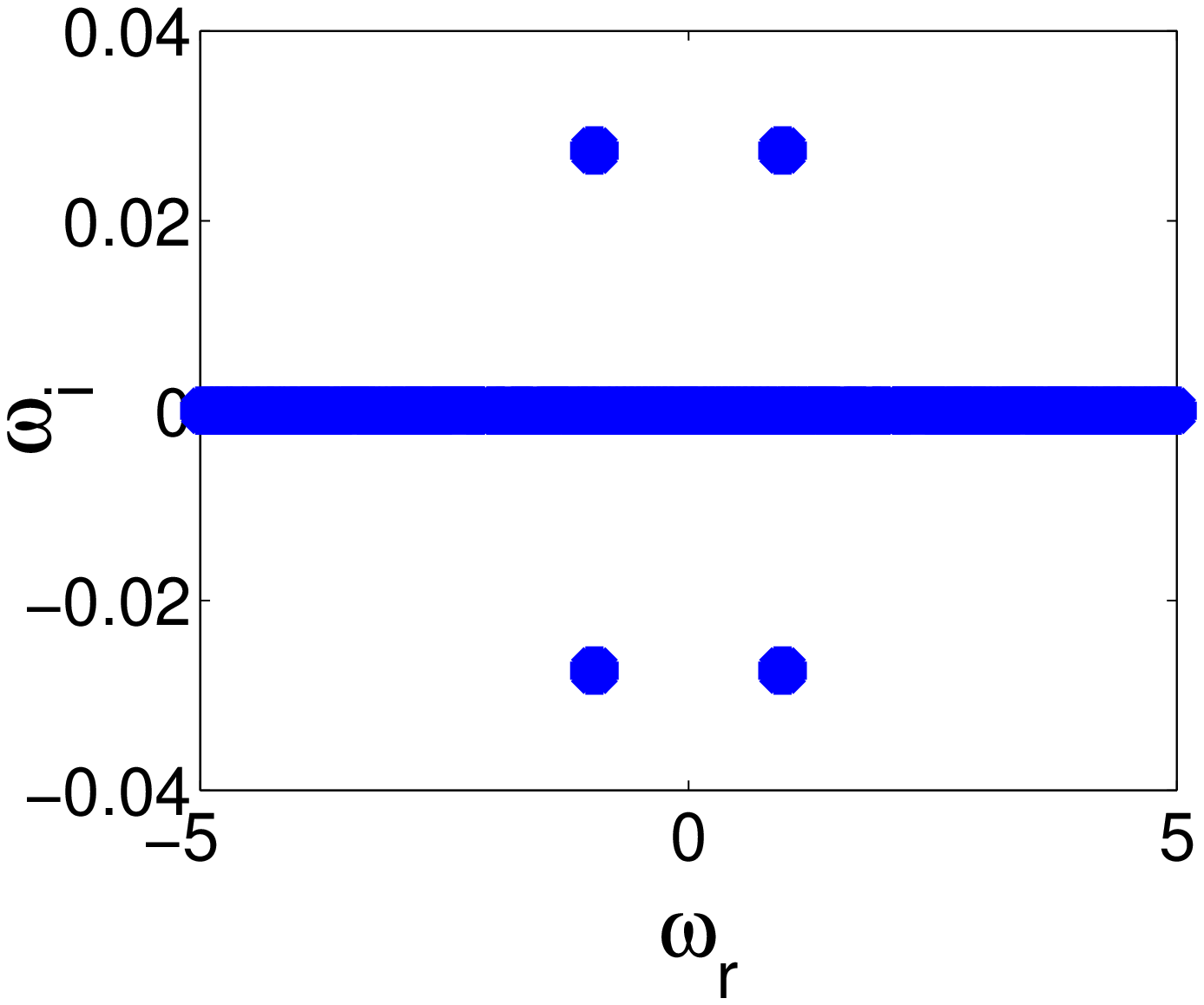}
\includegraphics[width=0.22\textwidth]{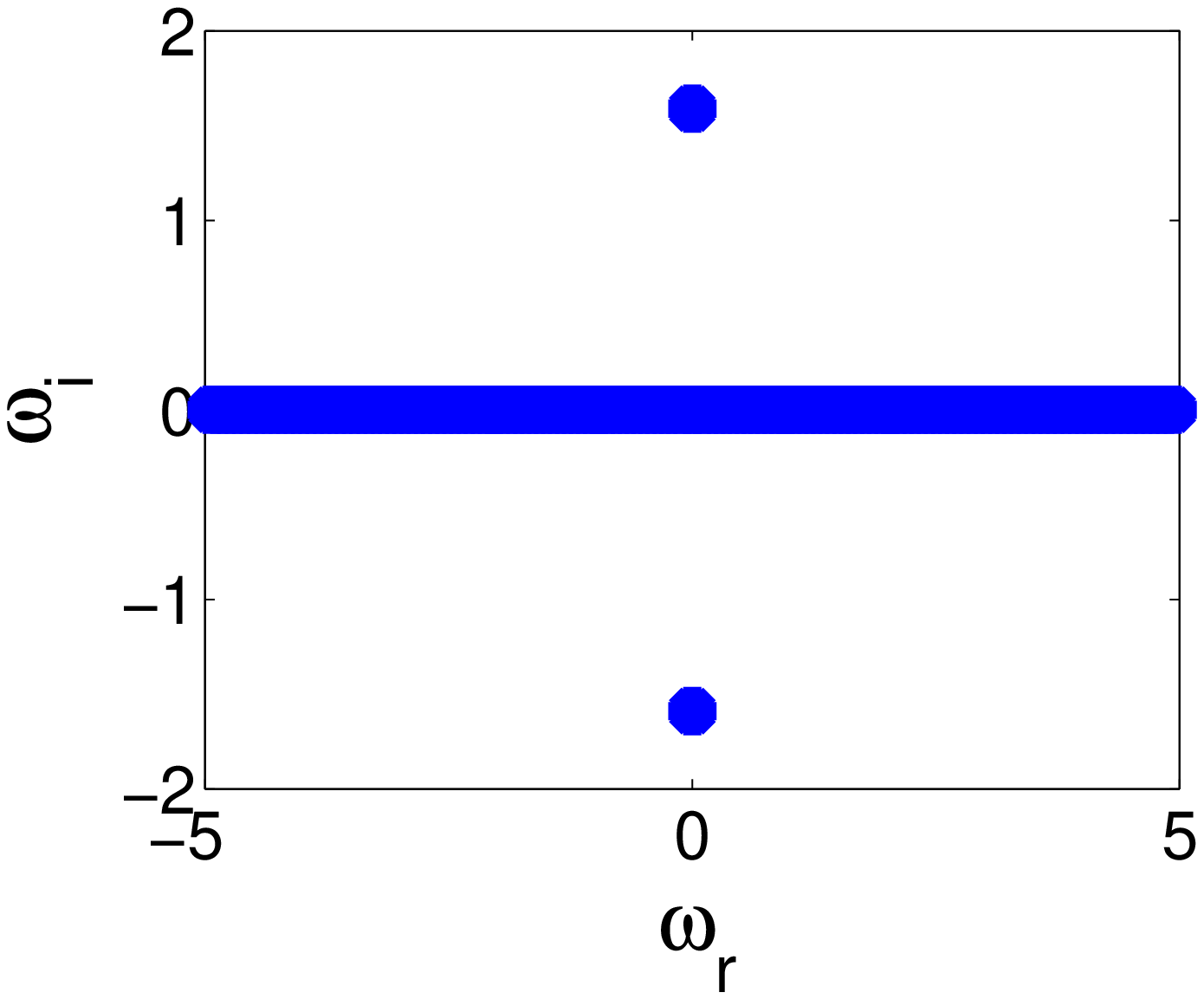} \\
\includegraphics[width=0.22\textwidth]{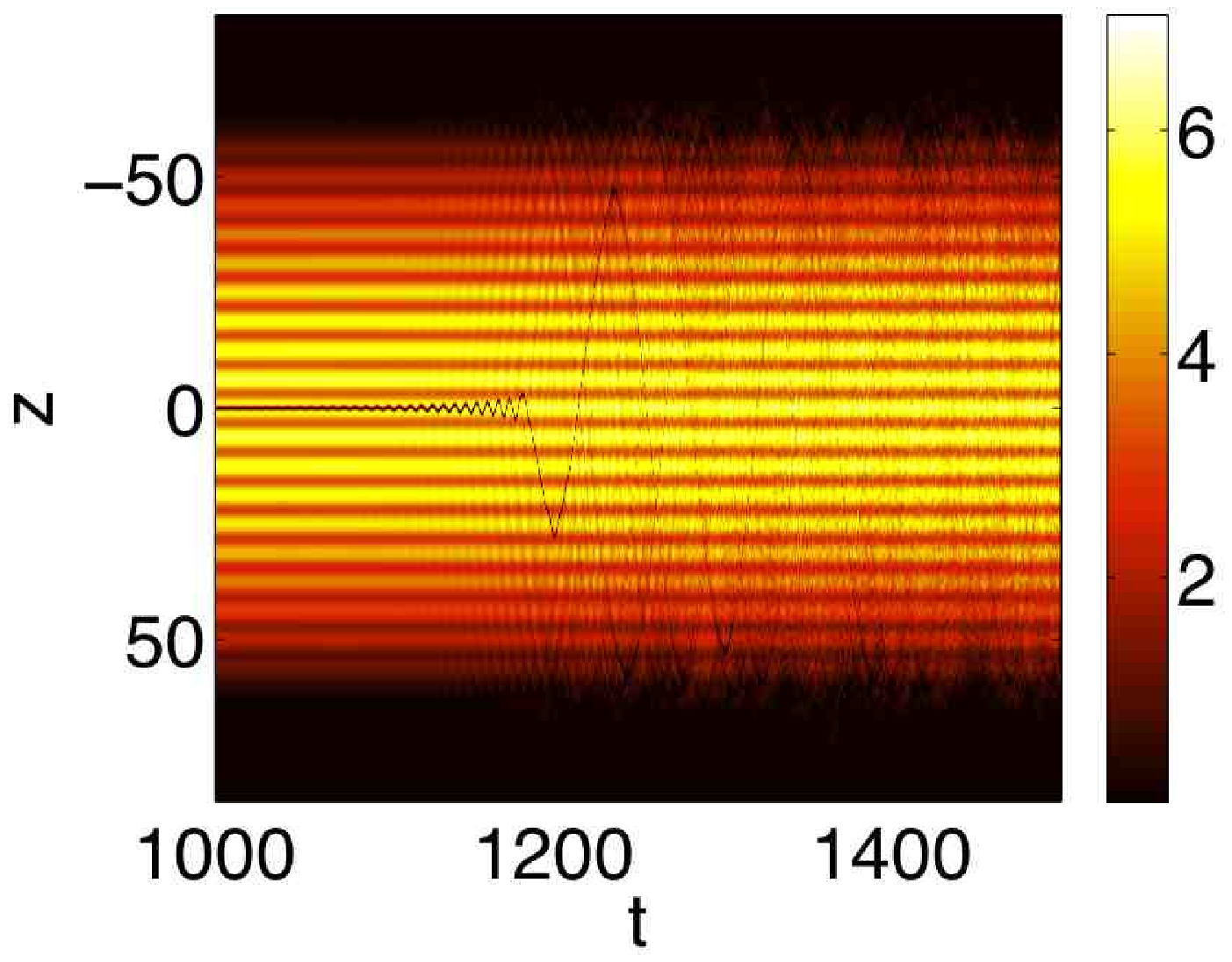}
\includegraphics[width=0.22\textwidth]{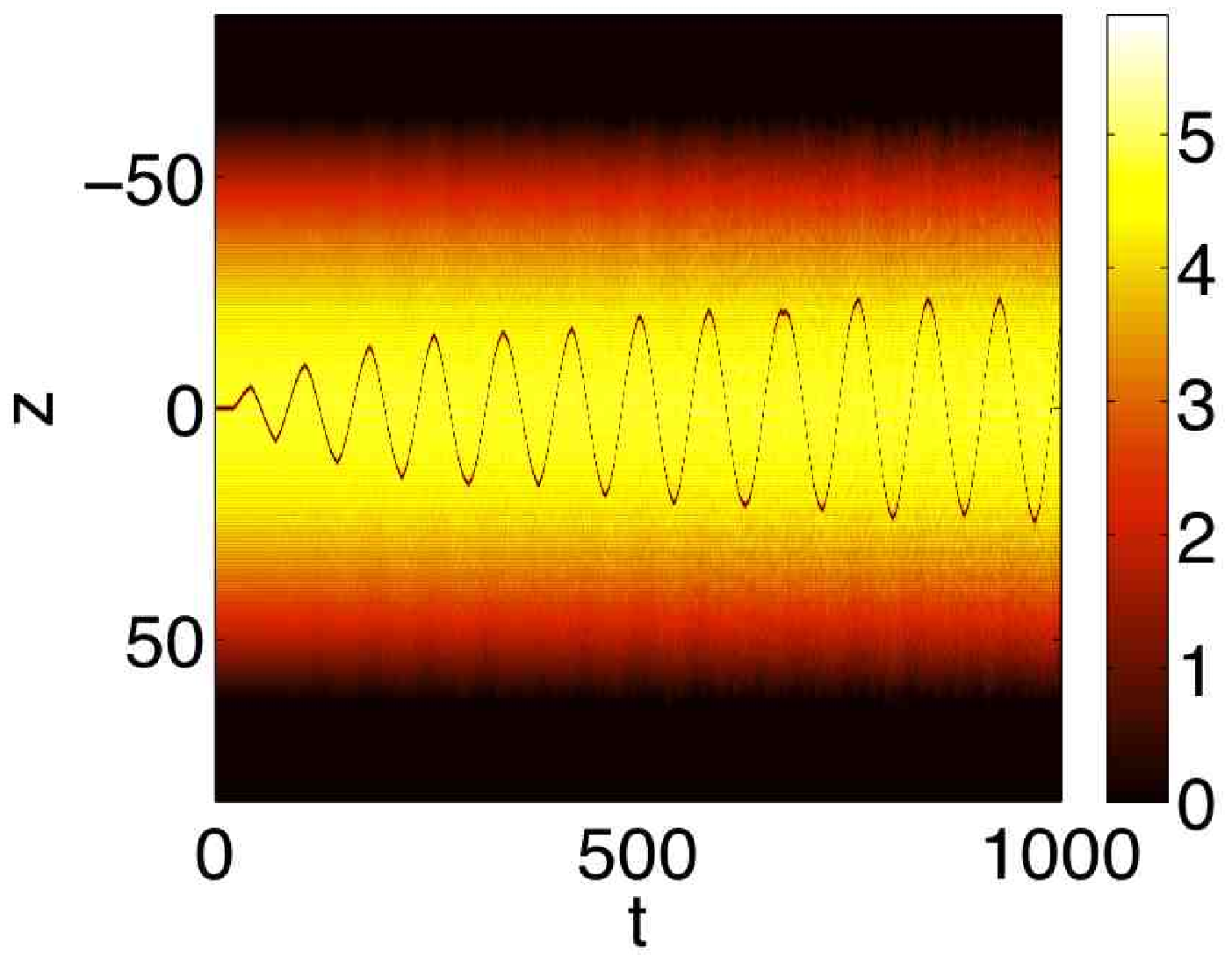}
\caption{[Color online] Dark solitary-wave solutions of the stationary GP equation with an aligned nonlinear lattice (top panels), their corresponding eigenfrequencies determined using the BdG equations (middle panels), and the spatio-temporal evolution of the solitary waves using simulations of the time-dependent GP equation (bottom panels) in an aligned nonlinear lattice.  The left column shows an unstable dark solitary wave with a quartet of complex eigenfrequencies; the parameter values are $g_m=1$, $g_s=0.5$, $k=1$, and $\mu=10$. The right column shows an unstable dark solitary wave with a pair of purely imaginary eigenfrequencies; the parameter values are $g_m=1$, $g_s=0.5$, $k=8$, and $\mu=10$.  In the top panels, we also show the harmonic trap (black dashed curves) and the inhomogeneity $\Delta g(z)$ (red dash-dotted curves).  We illustrate the spatio-temporal evolution of the BEC in the bottom panels; the color map indicates the value of $|u(z,t)|^2$.
} \label{numerics_ds}
\end{figure}

\begin{figure}[ht]
\centering
\includegraphics[width=0.22\textwidth]{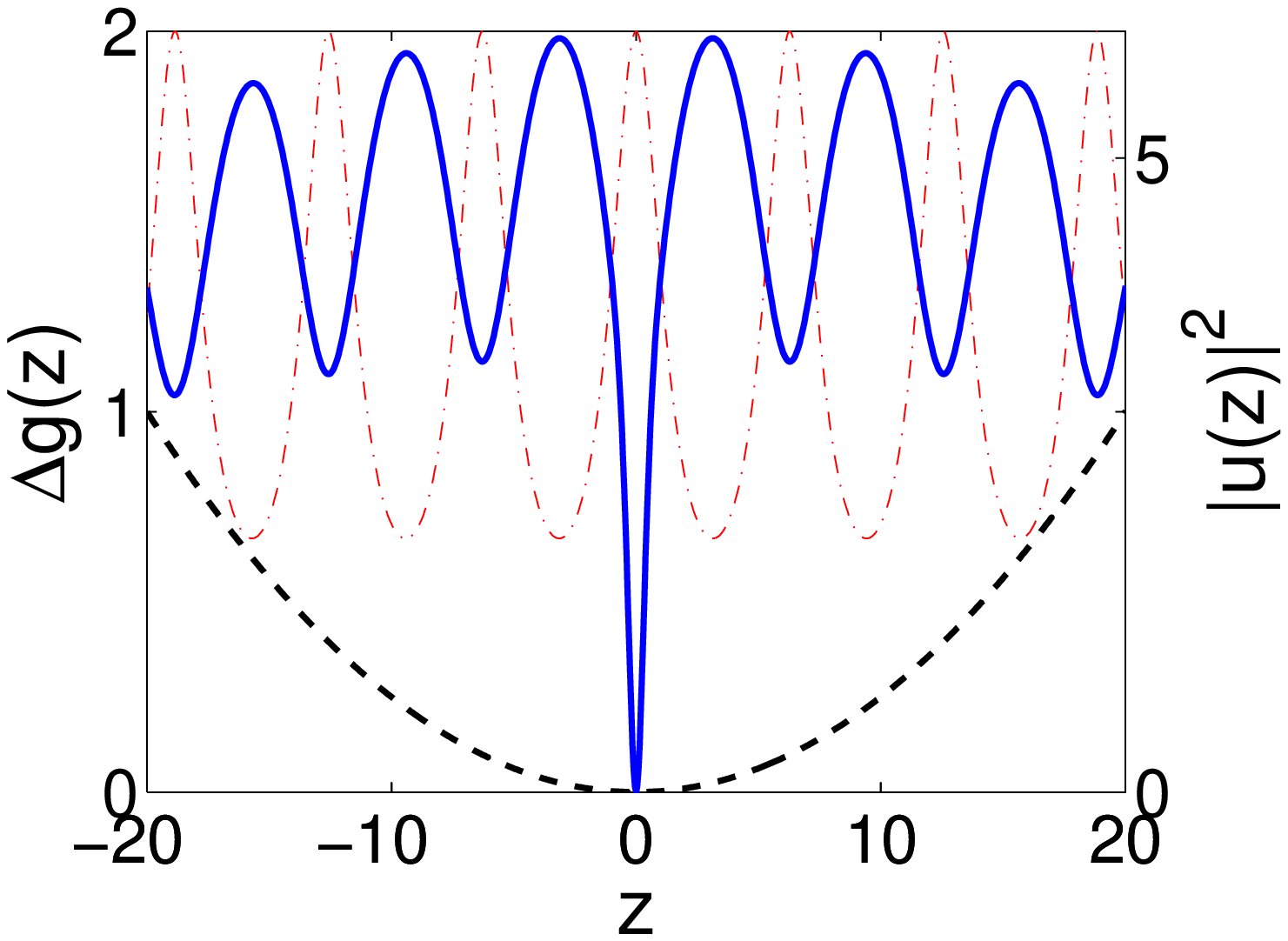}
\includegraphics[width=0.22\textwidth]{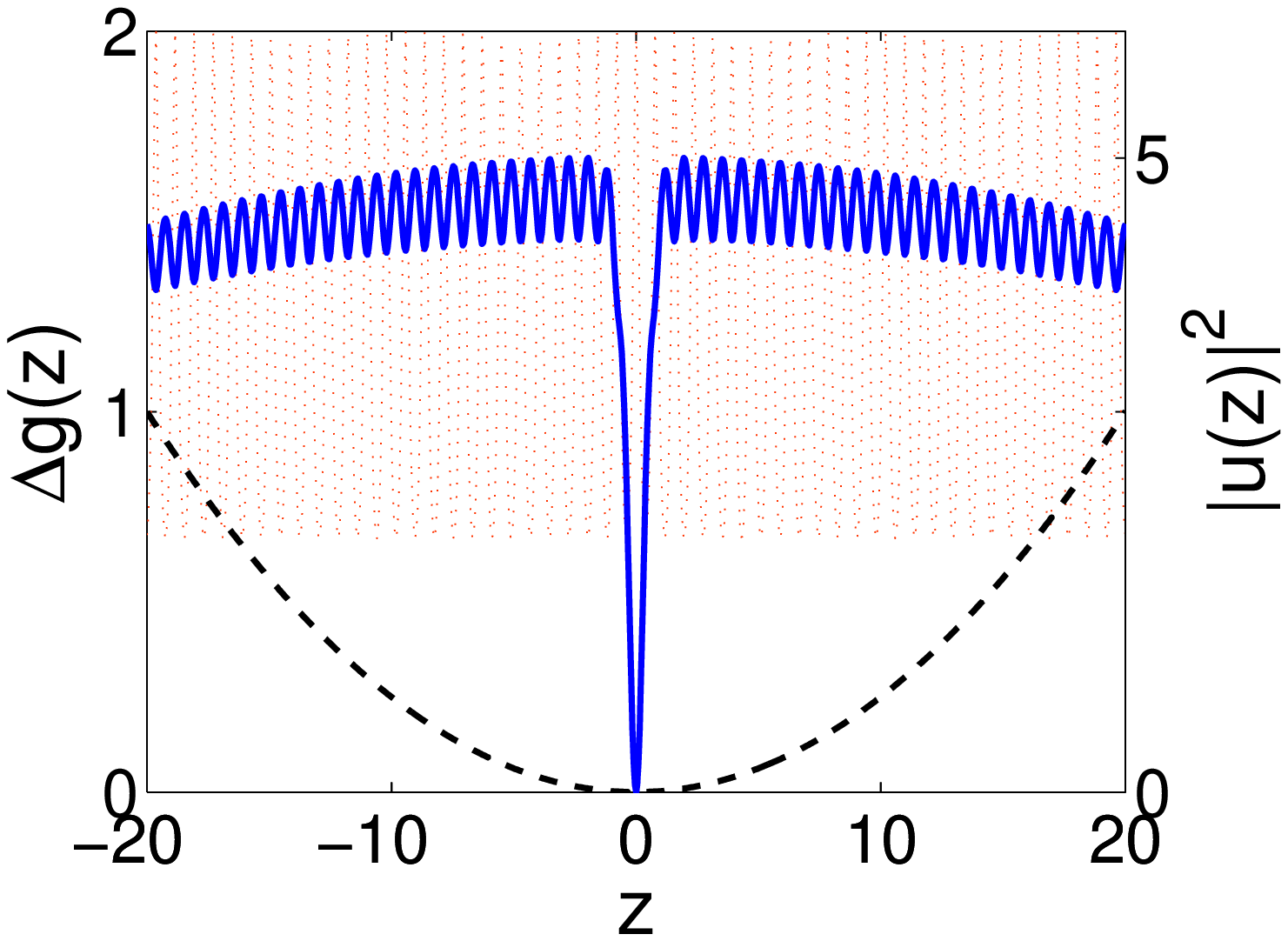}\\
\includegraphics[width=0.22\textwidth]{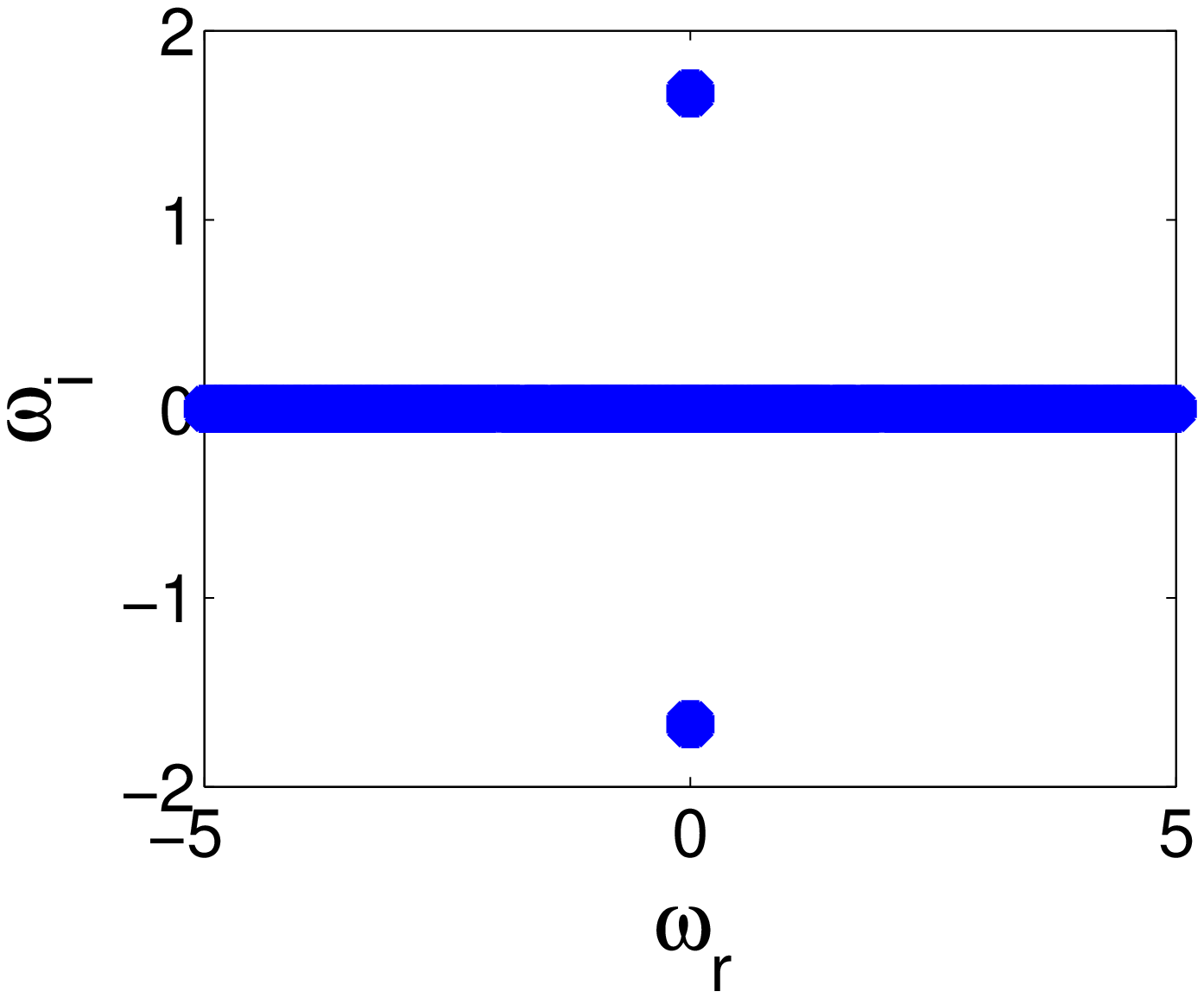}
\includegraphics[width=0.22\textwidth]{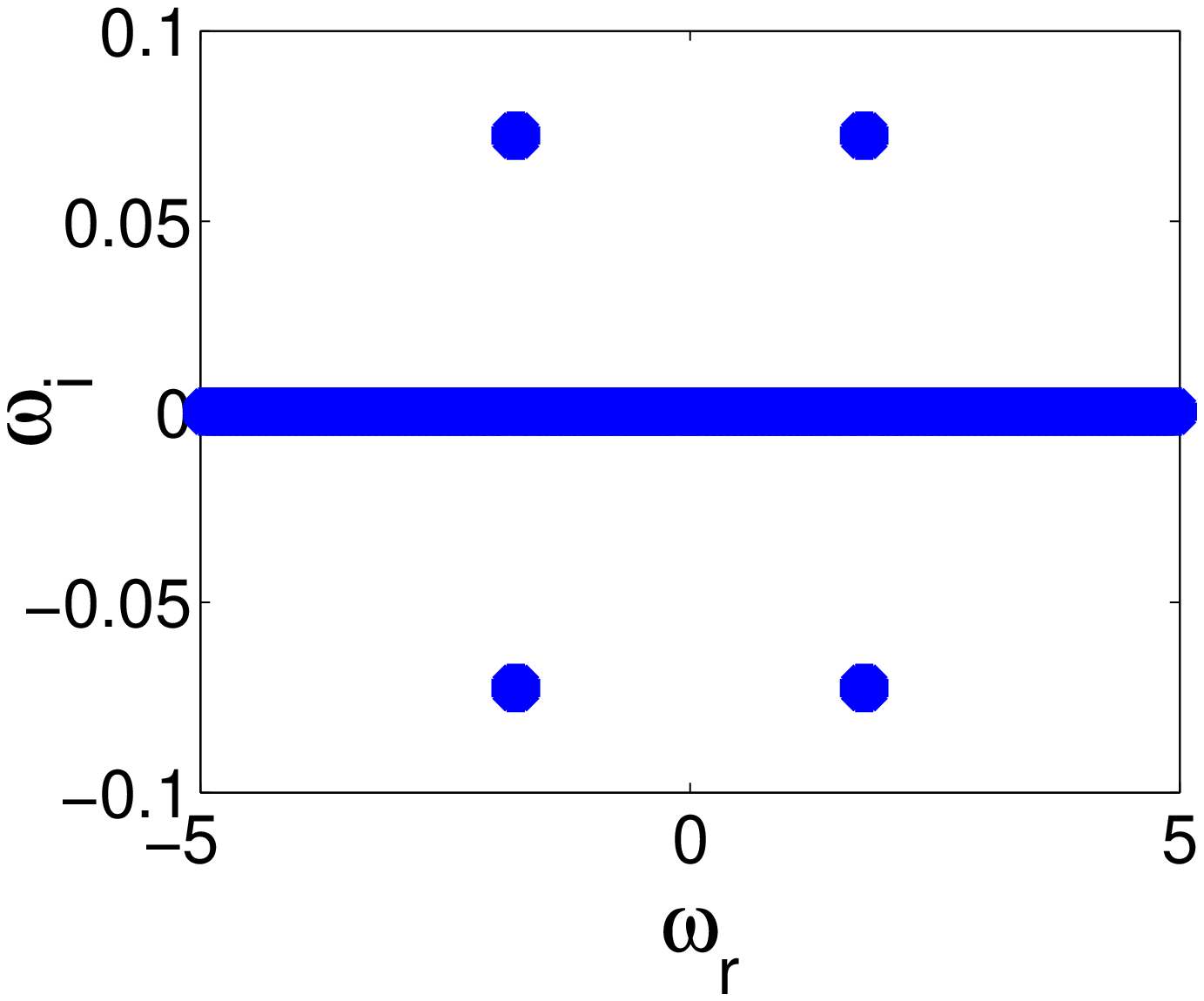}
\includegraphics[width=0.22\textwidth]{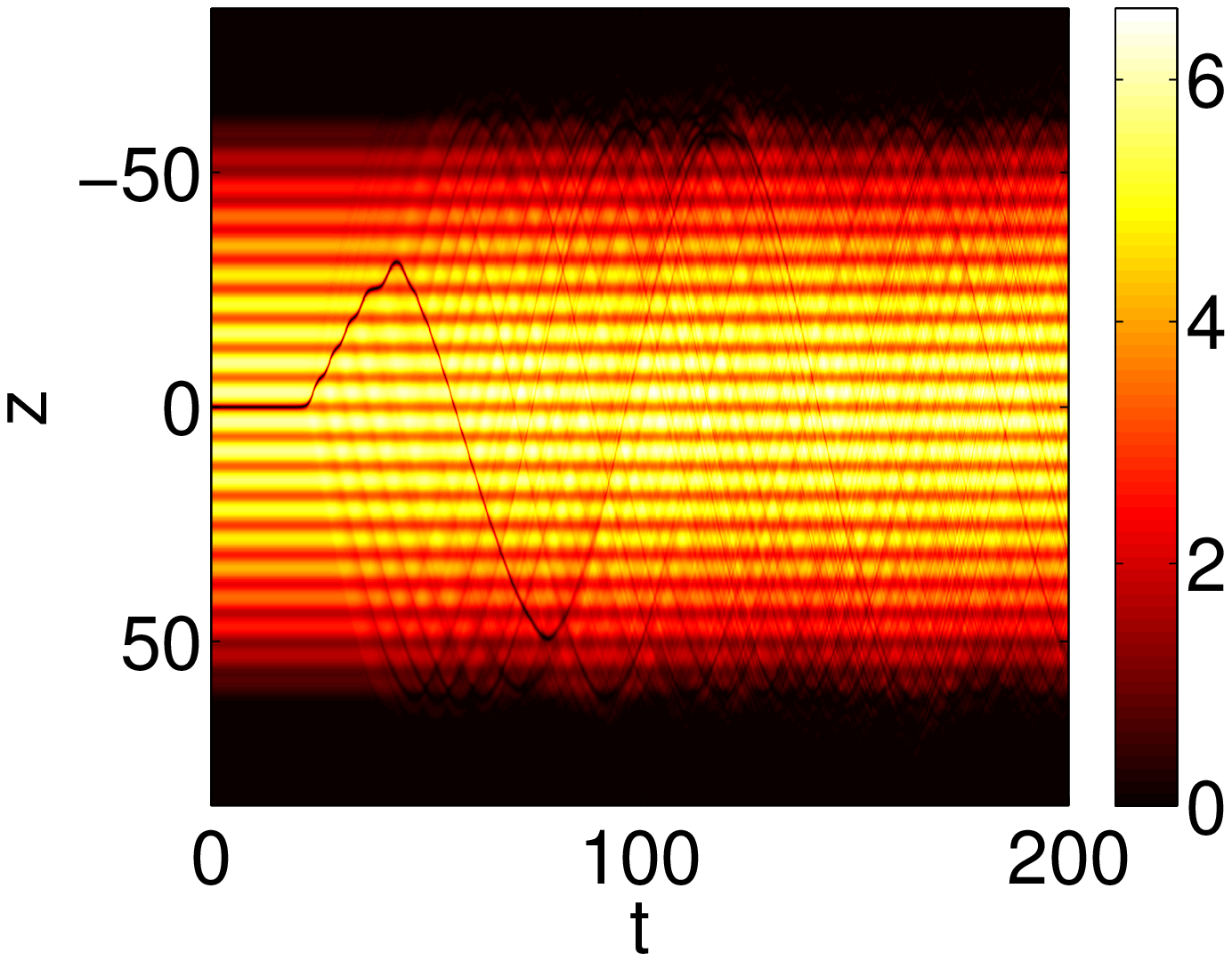}
\includegraphics[width=0.22\textwidth]{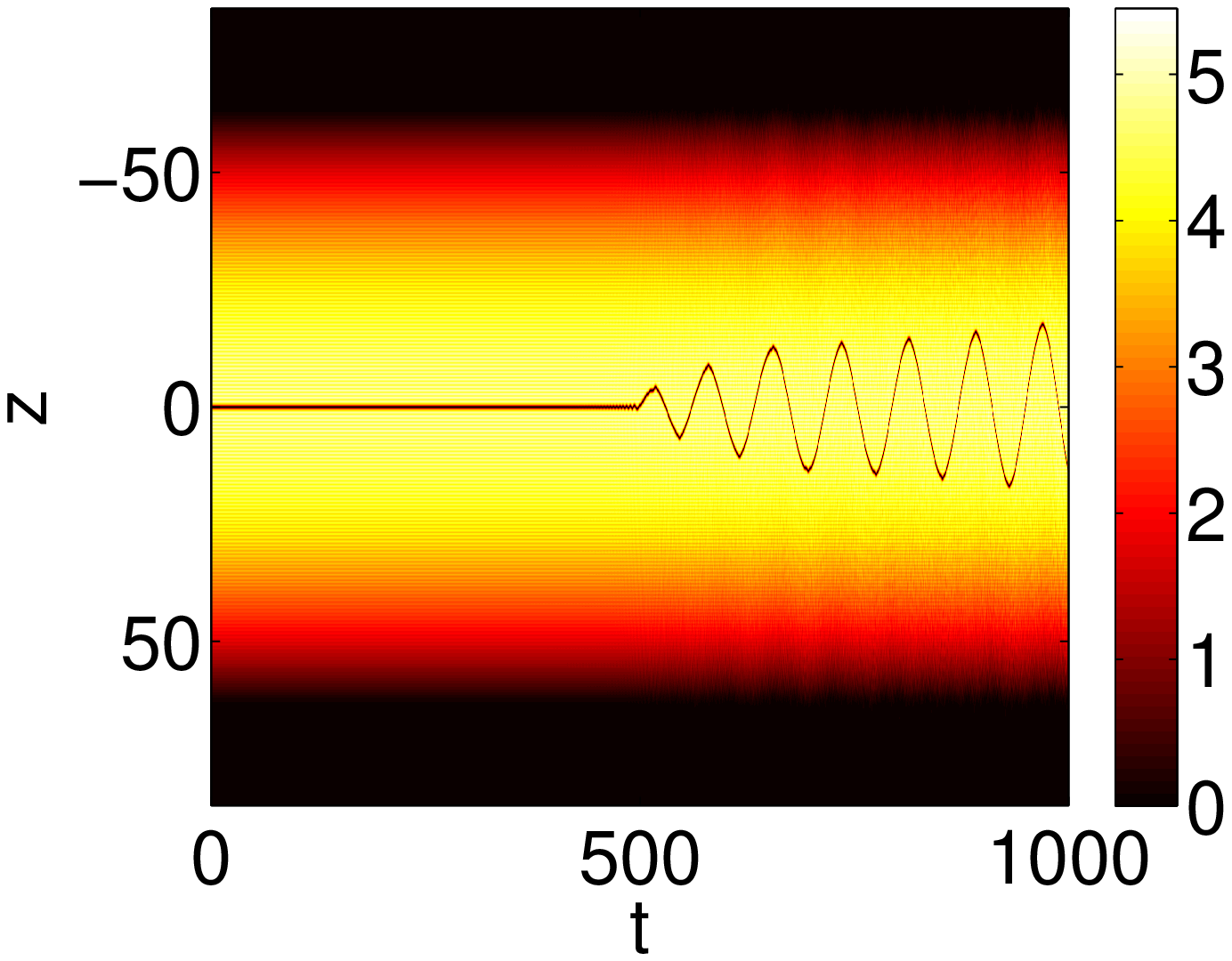}
\caption {[Color online] Dark solitary-wave solutions of the stationary GP equation (top panels), their corresponding eigenfrequencies determined using the BdG equations (middle panels), and the spatio-temporal evolution of the solitary waves using simulations of the time-dependent GP equation (bottom panels) in an anti-aligned nonlinear lattice. The left column shows unstable solitary waves with a pair of purely imaginary eigenfrequencies; the parameter values are $g_m=1$, $g_s=0.5$, $k=1$, and $\mu=10$.  The right column shows an unstable dark solitary wave with a quartet of complex eigenfrequencies; the parameter values are $g_m=1$, $g_s=0.5$, $k=8$, and $\mu=10$. In the top panels, we also show the harmonic trap (black dashed curves) and the inhomogeneity $\Delta g(z)$ (red dash-dotted curves).  We illustrate the spatio-temporal evolution of the BEC in the bottom panels; the color map indicates the value of $|u(z,t)|^2$.
}
\label{numerics_ds2}
\end{figure}

As we will discuss in this section, the stability of the observed dark solitary waves depends on the parameter values of the nonlinear lattice.  In Figs.~\ref{numerics_ds} and \ref{numerics_ds2}, we show the dynamics for several situations in aligned and anti-aligned lattices.  We show solitary-wave solutions of the stationary GP equation (obtained via Newton iteration) in the top panels, where we also display the nonlinear lattice and the harmonic trap.  The dip in the wavefunctions near $z = 0$ helps illustrate that these are in fact dark solitary waves. We show their corresponding eigenfrequencies (computed using the BdG equations) in the middle panels, and we show the temporal evolution of the solitary waves using direct numerical simulations of the time-dependent GP equation (\ref{eq:basicGP}) in the bottom panels.

We observed two qualitatively distinct situations for both aligned and anti-aligned nonlinear lattices: unstable dark solitary waves with one pair of purely 
imaginary eigenfrequencies and unstable dark solitary waves with a quartet of complex eigenfrequencies.  As the value of the wavenumber $k$ increases, the solution in the aligned lattice undergoes a transition from windows of mild oscillatory instabilities, which are indicated by quartets of complex eigenfrequencies that result from Hamiltonian-Hopf bifurcations, to strong instabilities, which are indicated by purely imaginary frequencies that result from crossing the origin of the spectral plane.  Interestingly, we observe the opposite dependence on $k$ in the anti-aligned lattice.  We illustrate these small-$k$ (left panels) and large-$k$ (right panels) results for aligned lattices in Fig.~\ref{numerics_ds} and for anti-aligned lattices in Fig.~\ref{numerics_ds2}.  

The dichotomy between the dynamics for aligned and anti-aligned lattices indicates that, given a fixed set of parameter values, one can in principle shift the 
nonlinear lattice (or, equivalently, the magnetic trap) and control the strength of the instability of a dark solitary wave.  This makes it possible to transition between a regime of strong instability and a regime of alternating windows of weak instability.  From the spatio-temporal evolution of the dark solitary waves, which we illustrate in the bottom panels, we observe in both cases that the waves ultimately become displaced from the center and oscillate between two turning points in the parabolically trapped BEC.  In the small-$k$ regime, the strong interaction of the propagating dark solitary-wave structures leads to their eventual decay after only a few oscillations. When $k$ is large, however, the BEC appears to reach a ``homogenized'' limit that is characterized by a reduced range of spatial variation (due to the nonlinear lattice) within the trap. Ultimately, as $k$ increases further, the role of the nonlinear lattice is to yield an effective averaged modulation of the $s$-wave scattering length.

\begin{figure} [h!t]
\centering 
\includegraphics[width=.5\textwidth]{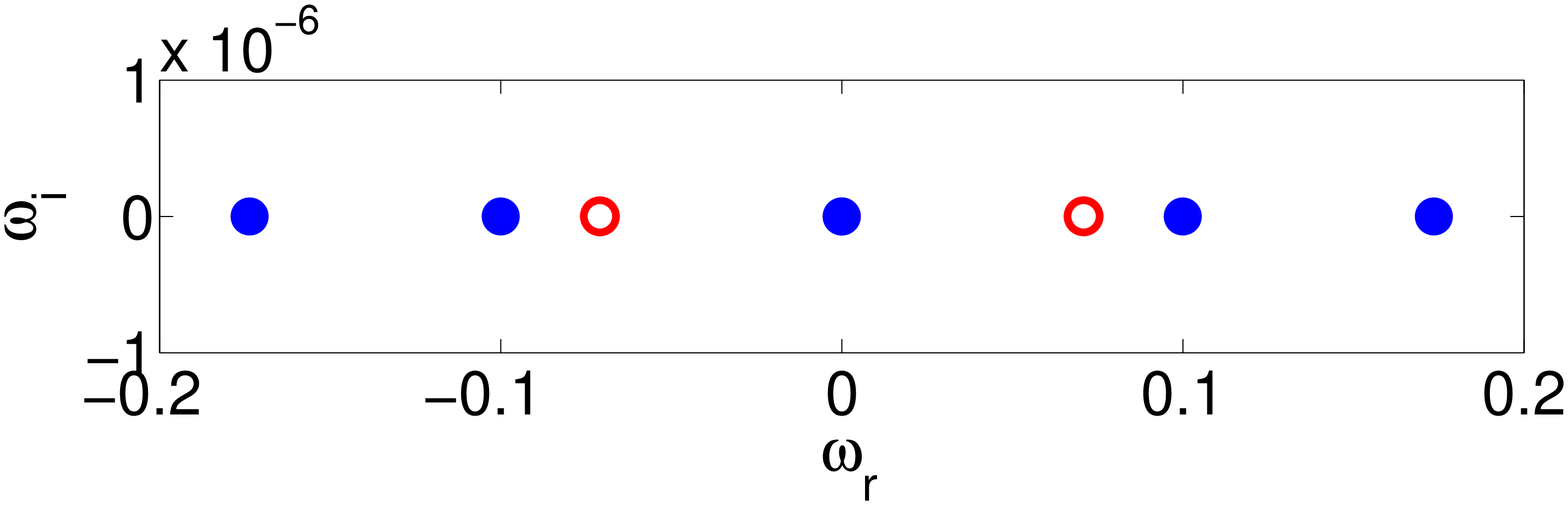}
\includegraphics[width=.5\textwidth]{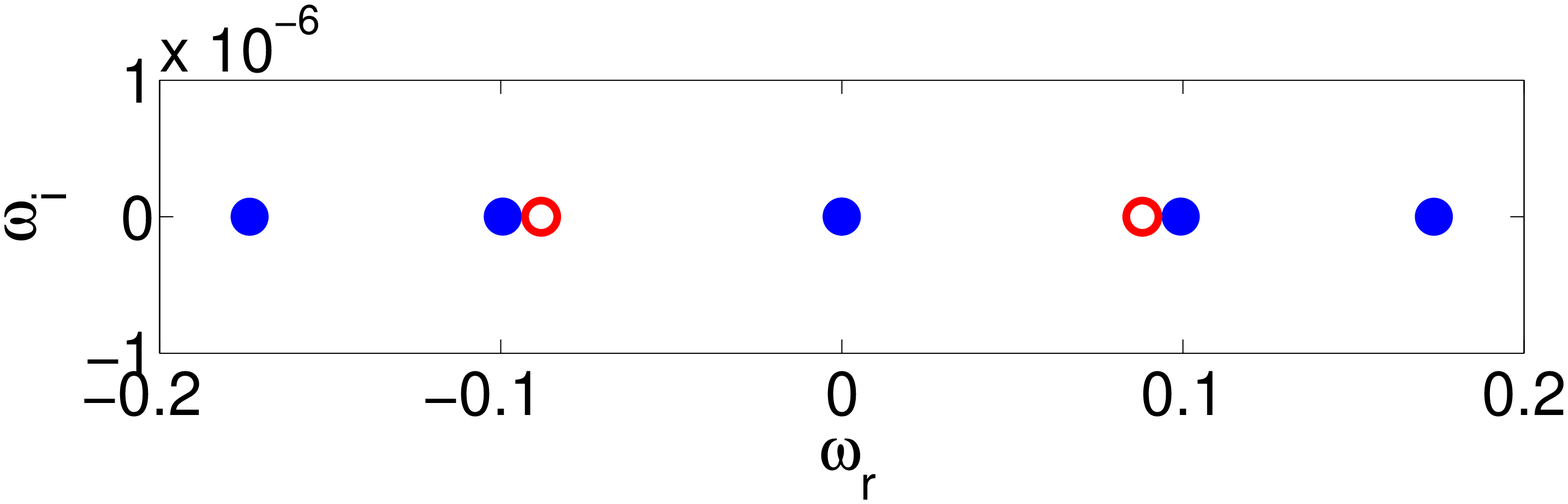}
\includegraphics[width=.5\textwidth]{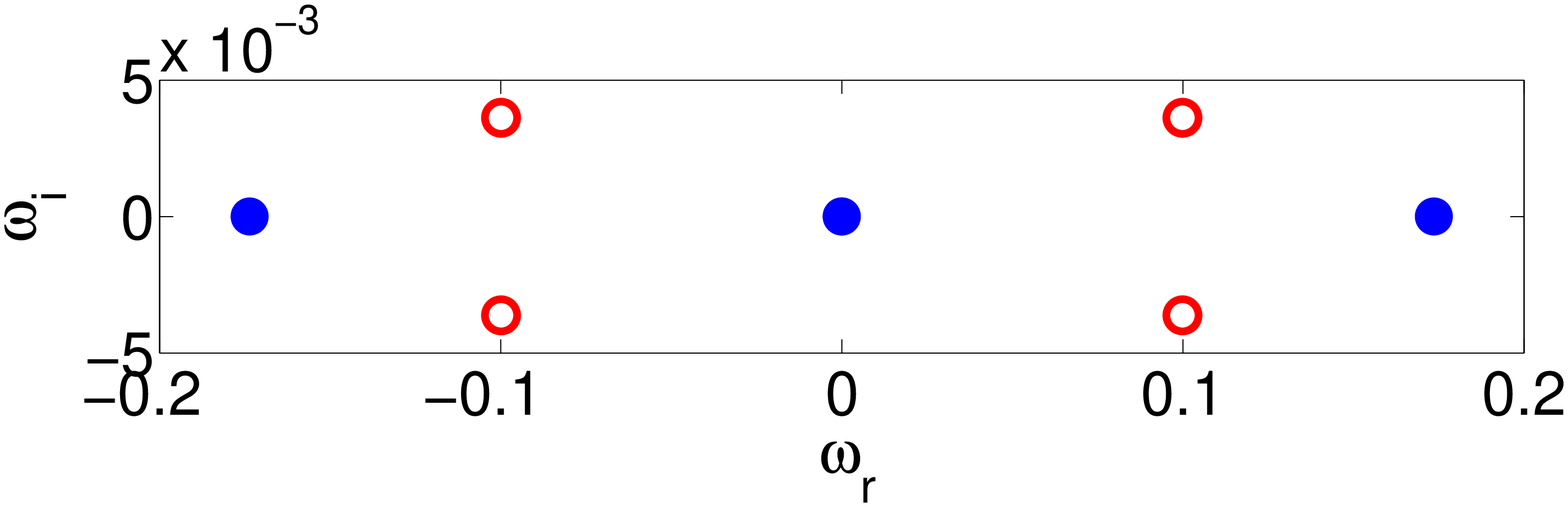}
\includegraphics[width=.5\textwidth]{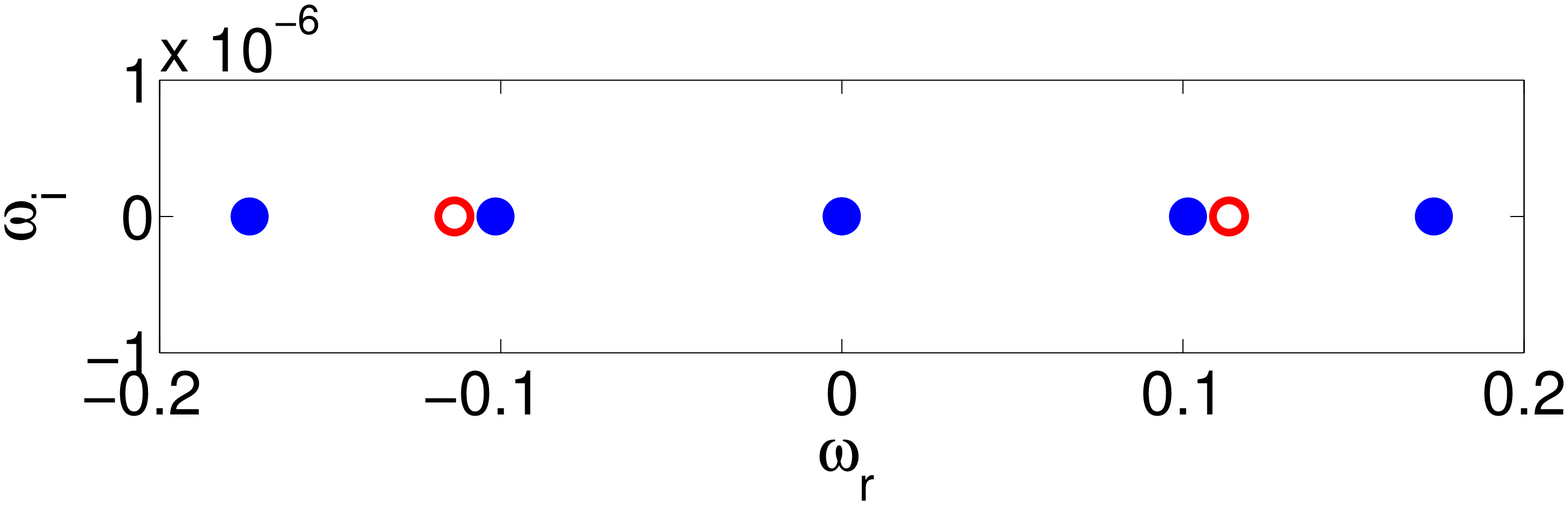}
\caption{[Color online] Variation of the eigenfrequency spectrum as $g_m$ increases from $0$ for aligned lattices with $g_s=0.5$, $k=1$, and $\mu=10$.  The anomalous mode, which we plot using red circles, moves to larger frequencies as $g_m$ increases.  From top to bottom, the values of $g_m$ are $0$, $0.005$, $0.008$, and $0.015$.} \label{align_mode_k1}
\end{figure}

\begin{figure} [h!t]
\centering 
\includegraphics[width=0.5\textwidth]{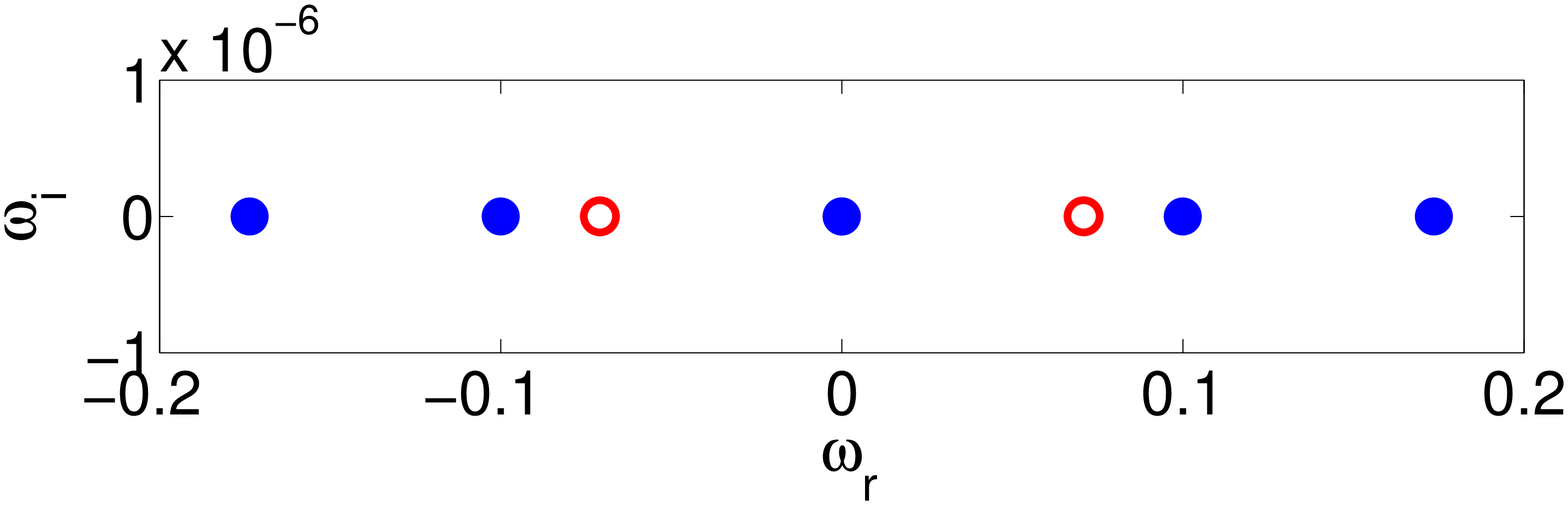}
\includegraphics[width=0.5\textwidth]{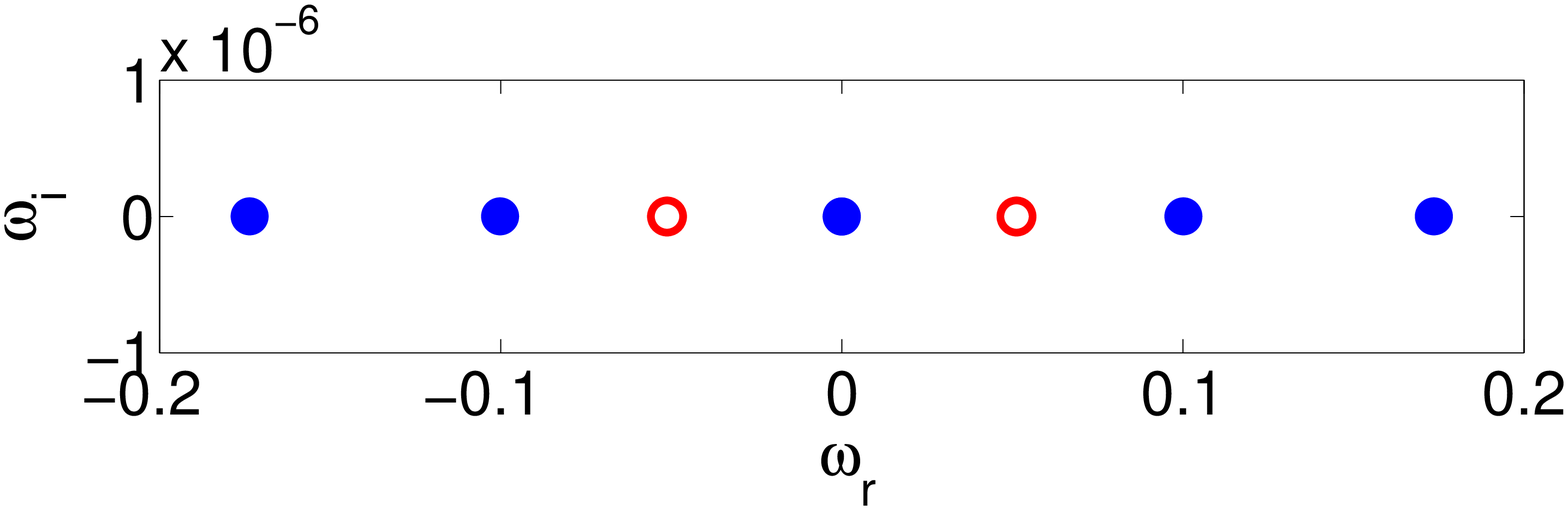}
\includegraphics[width=0.5\textwidth]{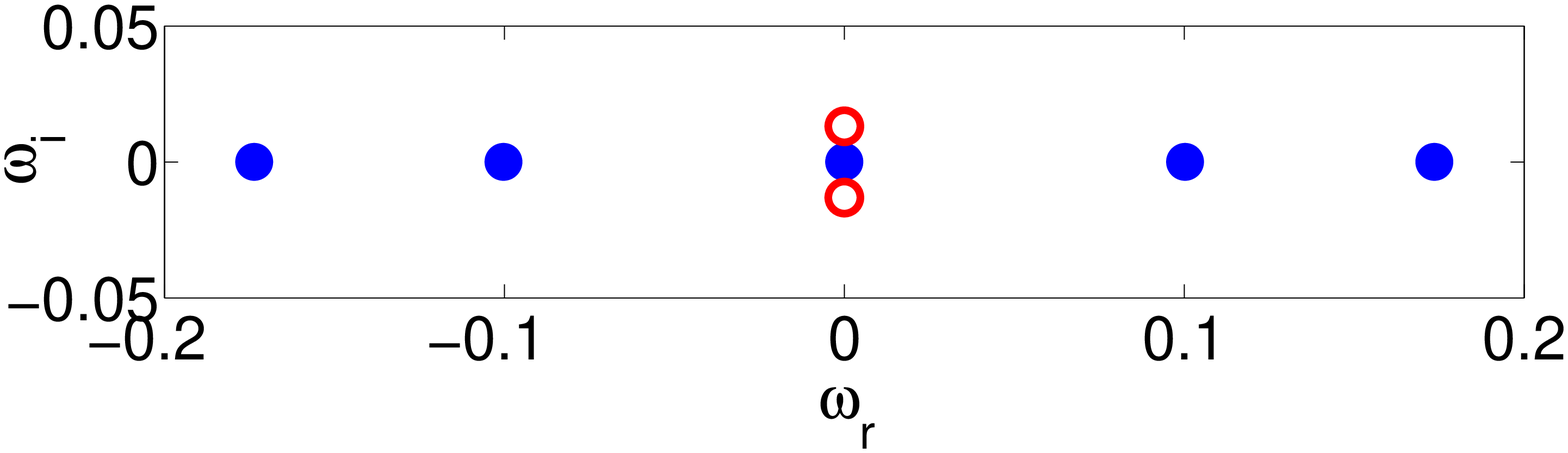}
\includegraphics[width=0.5\textwidth]{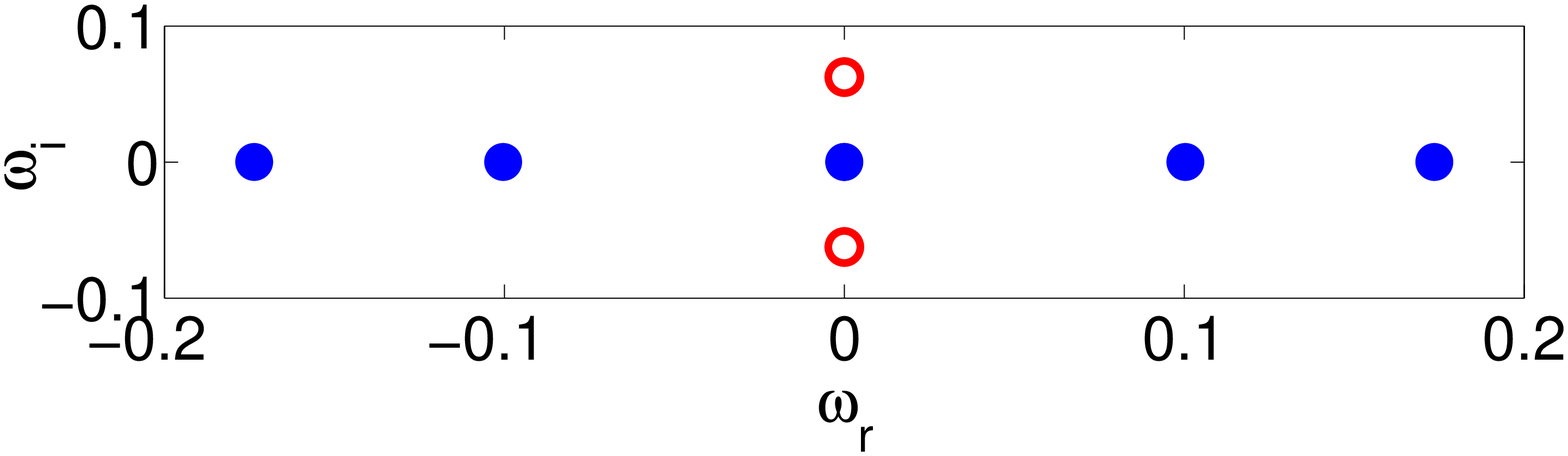}
\caption{[Color online] Variation of the eigenfrequency spectrum as $g_m$ increases from $0$ for aligned lattices with $g_s=0.5$, $k=8$, and $\mu=10$.  The anomalous mode, which we plot using red circles, moves to lower frequencies (which eventually become imaginary) as $g_m$ increases.  From top to bottom, the values of $g_m$ are $0$, $0.005$, $0.008$, and $0.015$.} \label{align_mode_k8}
\end{figure}

\begin{figure} [h!t]
\centering 
\includegraphics[width=0.5\textwidth]{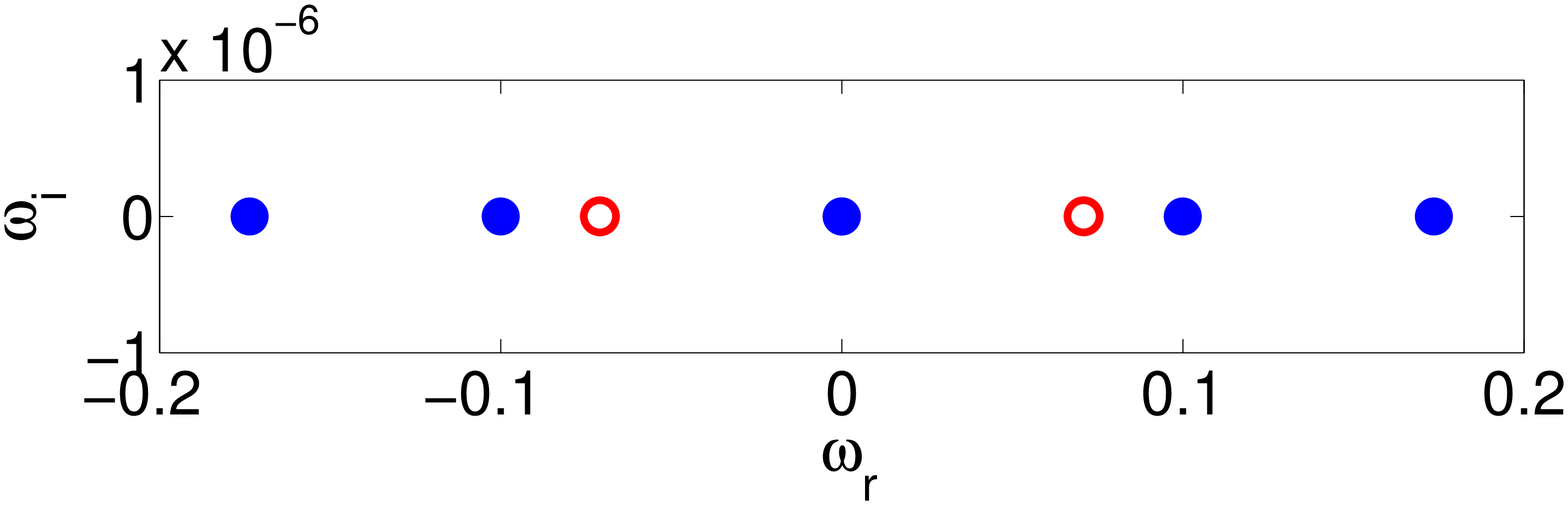}
\includegraphics[width=0.5\textwidth]{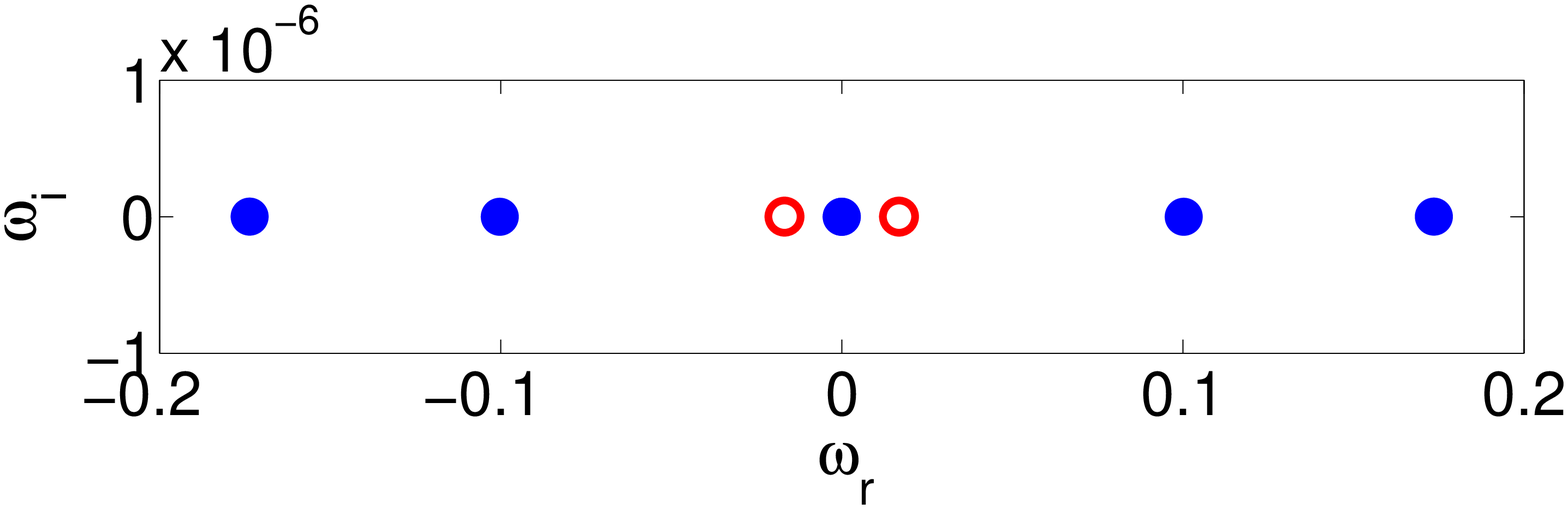}
\includegraphics[width=0.5\textwidth]{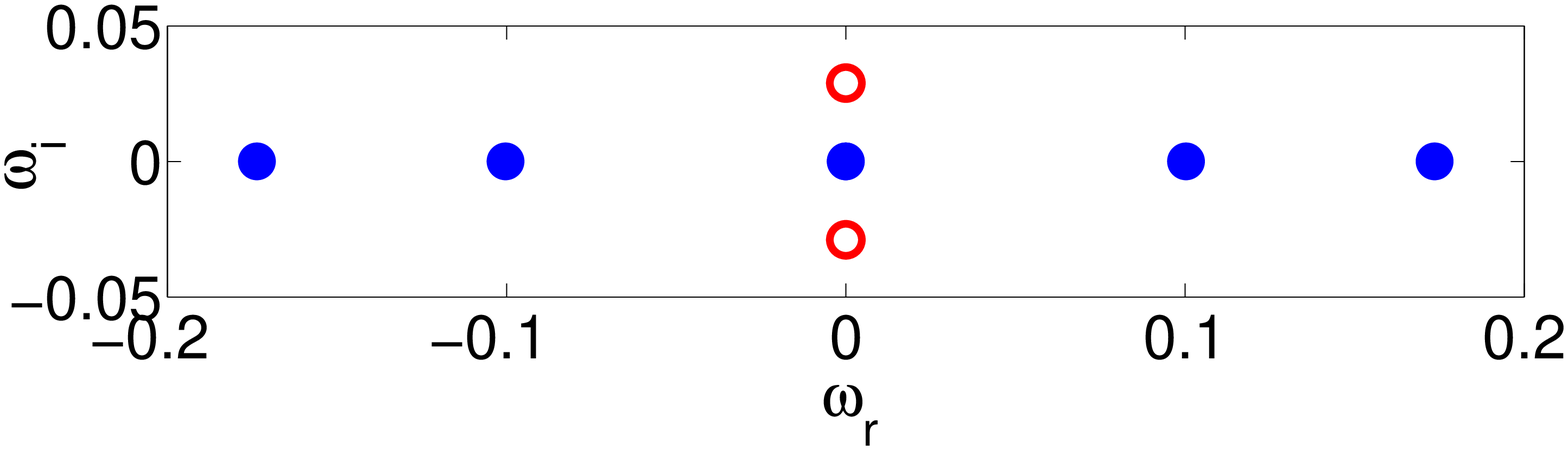}
\includegraphics[width=0.5\textwidth]{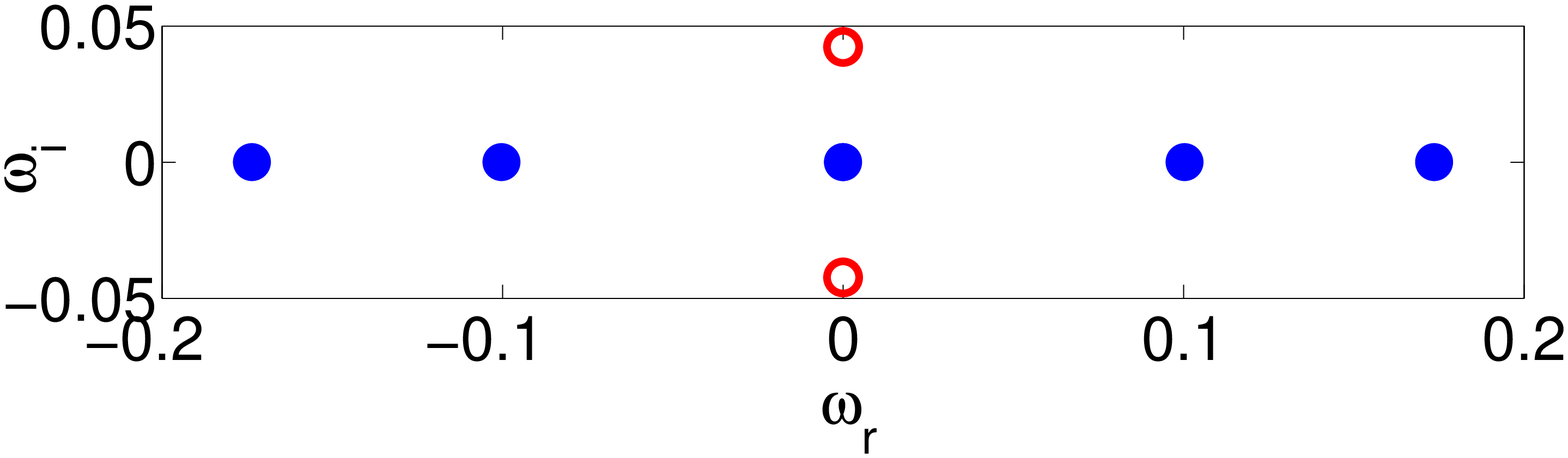}
\caption{[Color online] Variation of the eigenfrequency spectrum as $g_m$ increases from 
$0$ for anti-aligned lattices with $g_s=0.5$, $k=1$, and $\mu=10$.  The anomalous mode, which we plot using red circles, moves to lower frequencies (which eventually become imaginary) as $g_m$ increases.  From top to bottom, the values of $g_m$ are $0$, $0.007$, $0.008$, and $0.01$.}  \label{anti_mode_k1}
\end{figure}

\begin{figure} [h!t]
\centering 
\includegraphics[width=0.5\textwidth]{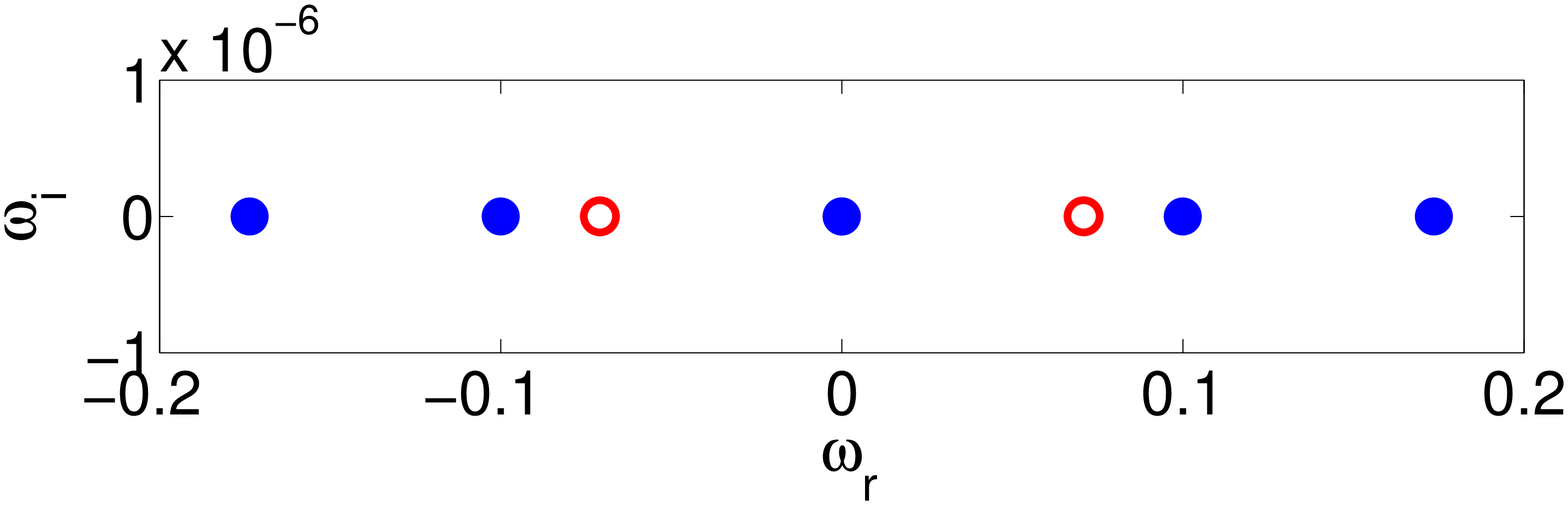}
\includegraphics[width=0.5\textwidth]{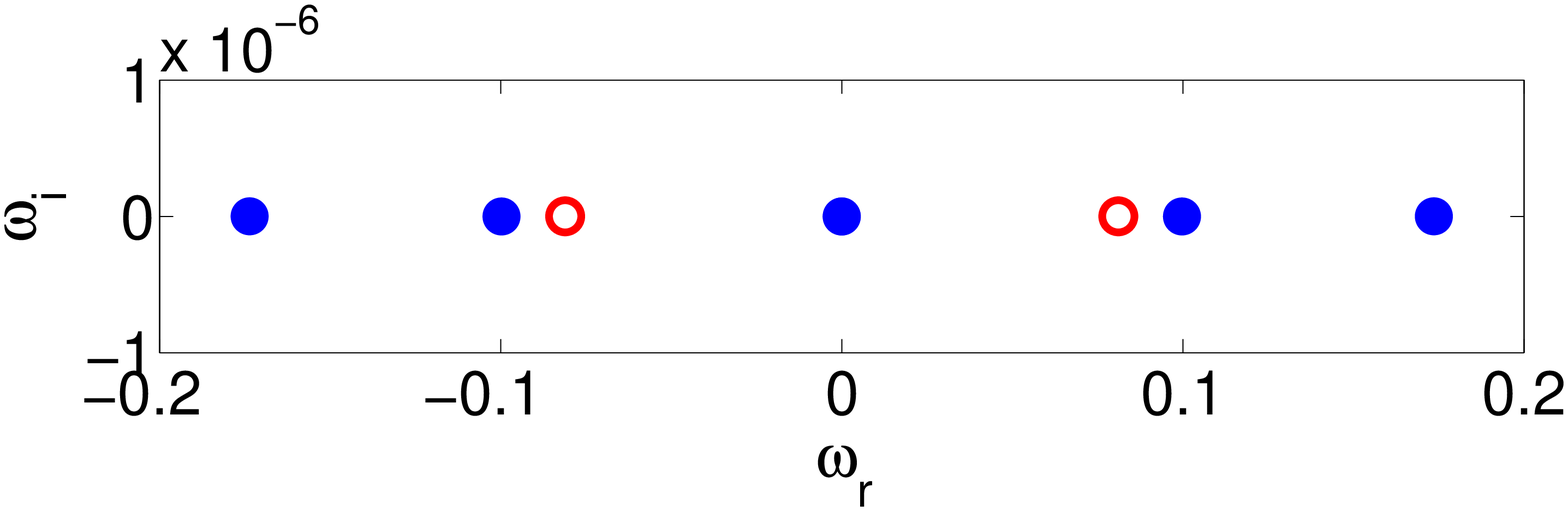}
\includegraphics[width=0.5\textwidth]{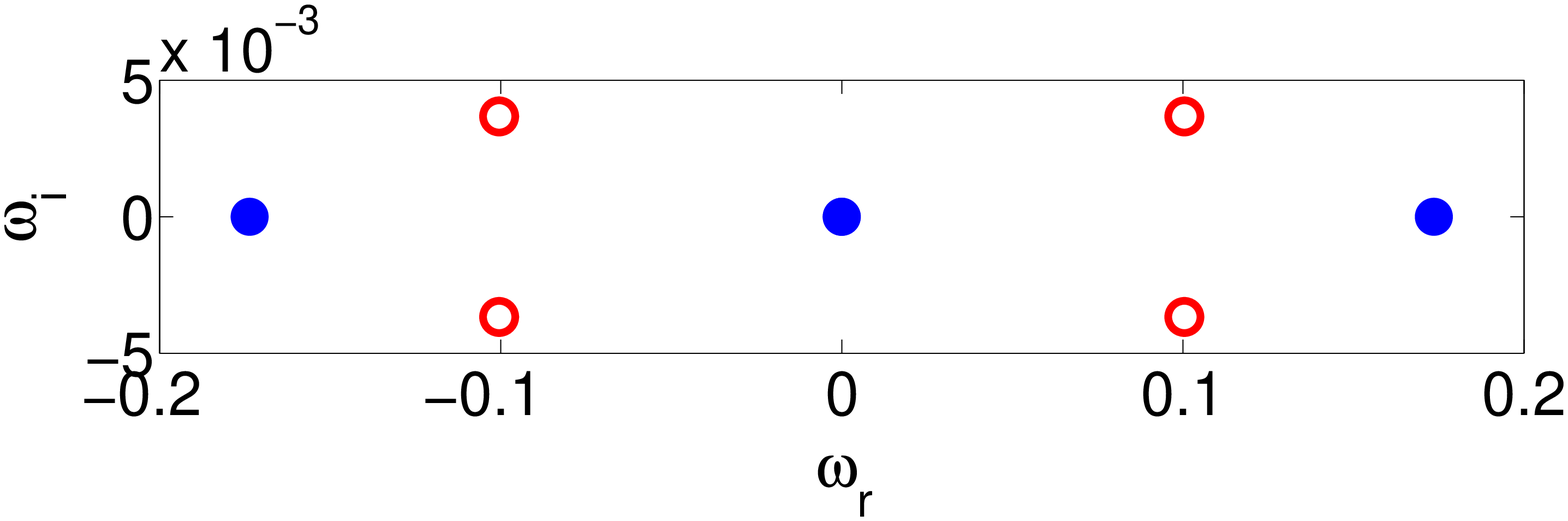}
\includegraphics[width=0.5\textwidth]{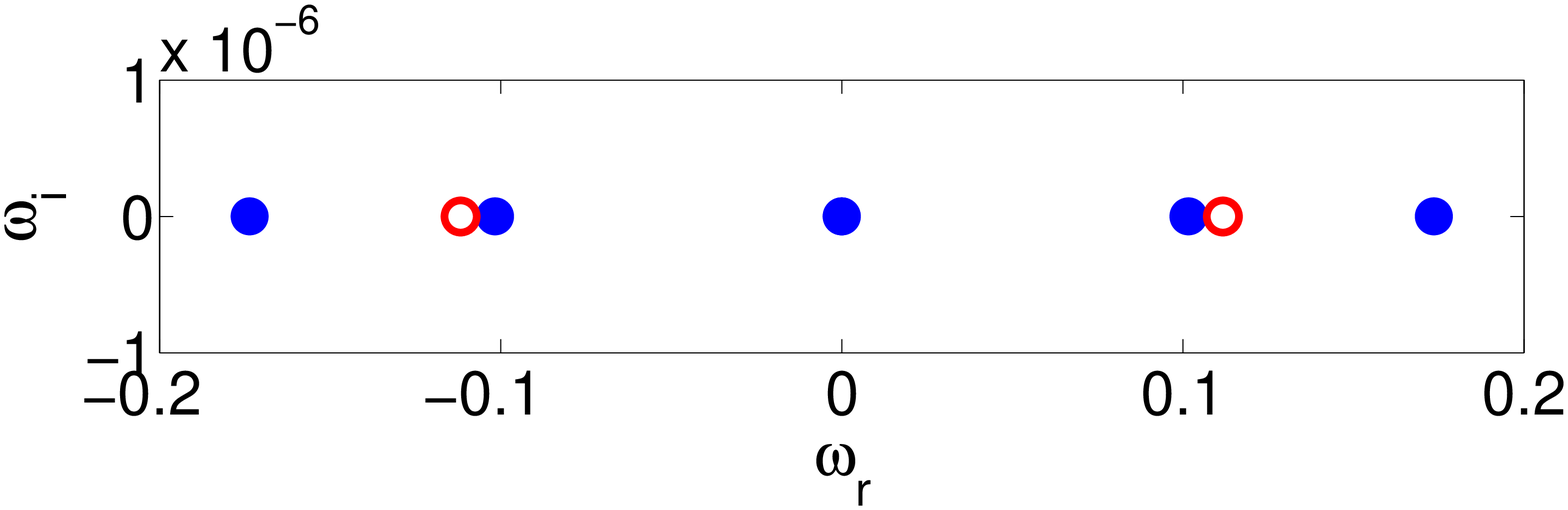}
\caption{[Color online] Variation of the eigenfrequency spectrum as $g_m$ increases from $0$ for anti-aligned lattices with $g_s=0.5$, $k=8$, and $\mu=10$.  The anomalous mode, which we plot using red circles, moves to higher frequencies as $g_m$ increases.  From top to bottom, the values of $g_m$ are $0$, $0.002$, $0.005$, and $0.007$.}  \label{anti_mode_k8}
\end{figure}

Given a regime of the wavenumber $k$, we vary $g_m$ from $0$ by a small amount and
monitor the change of eigenfrequency spectrum that we obtain using the 
BdG analysis.  In particular, we examine the four (pairs of) eigenfrequencies of smallest magnitude.  We illustrate typical trends of the eigenfrequency spectrum variation in
Figs.~\ref{align_mode_k1}--\ref{anti_mode_k8}.  In order, these figures show results for the small-$k$ regime for a BEC in an aligned nonlinear lattice, the large-$k$ regime for an aligned lattice, the small-$k$ regime for an anti-aligned lattice, and the large-$k$ regime for
an anti-aligned lattice.  In all eigenfrequency computations, we are concerned 
predominantly with the so-called anomalous mode (i.e., the
only mode that has 
negative Krein signature \cite{ds_dimitri}).
The anomalous mode is located at $\omega = \pm 1/(B \sqrt{2})$ for $g_m=0$, and we plot it using red circles in the figures.  In the small-$k$ regime of the aligned lattice and the large-$k$ regime of 
the anti-aligned lattice, the anomalous mode moves upward along the real axis as 
$g_m$ increases.  In these cases, its collision with another mode yields a Hamiltonian-Hopf bifurcation and an oscillatory instability.  In the 
large-$k$ regime of the aligned lattice and the small-$k$ regime of the 
anti-aligned lattice, the anomalous mode moves downward along the real axis as $g_m$ increases.  When it hits the origin of the spectral plane of eigenfrequencies, one obtains an exponential instability associated with an imaginary eigenfrequency.


\begin{figure} [h!t]
\centering 
\includegraphics[width=0.5\textwidth]{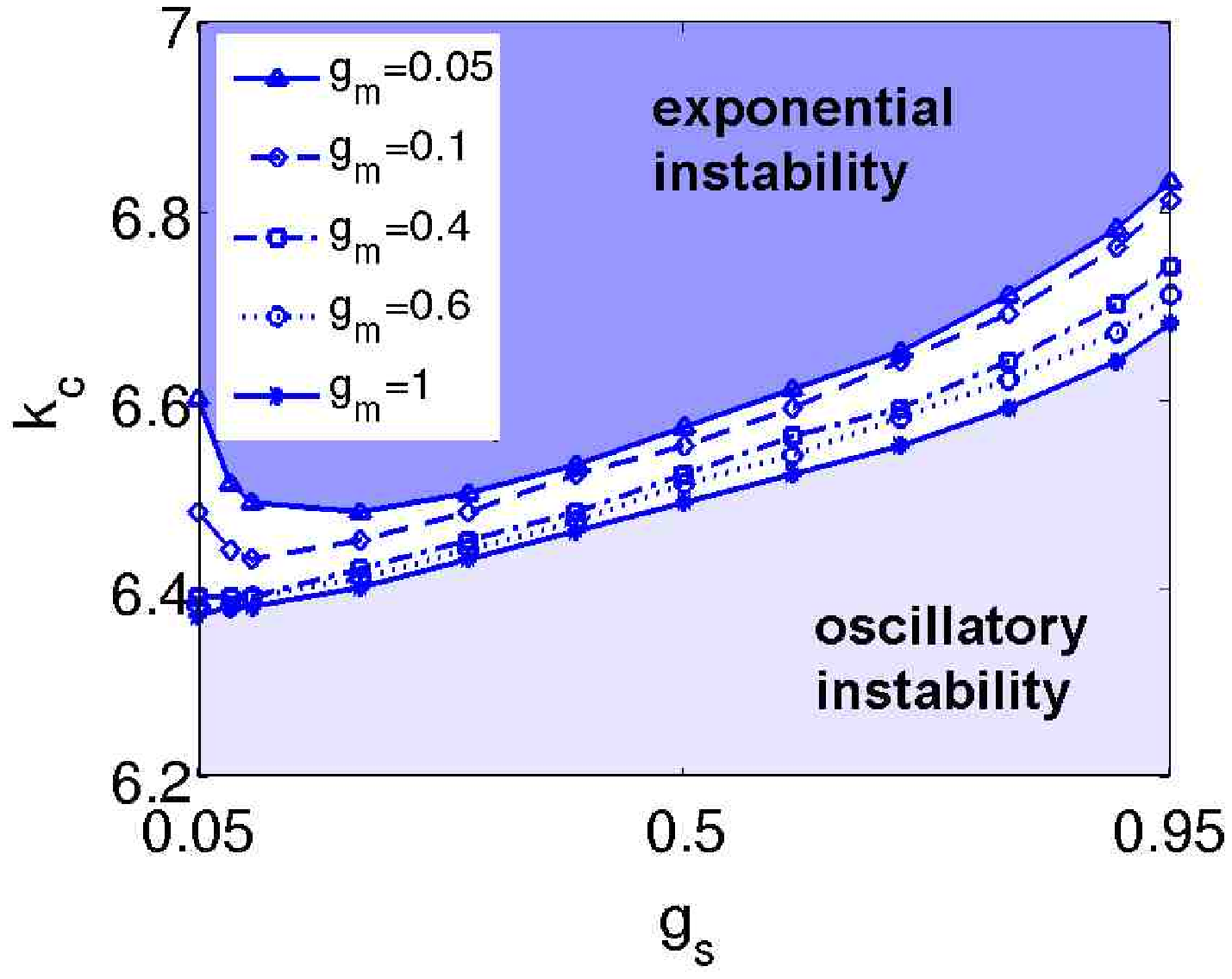}\\
\includegraphics[width=0.5\textwidth]{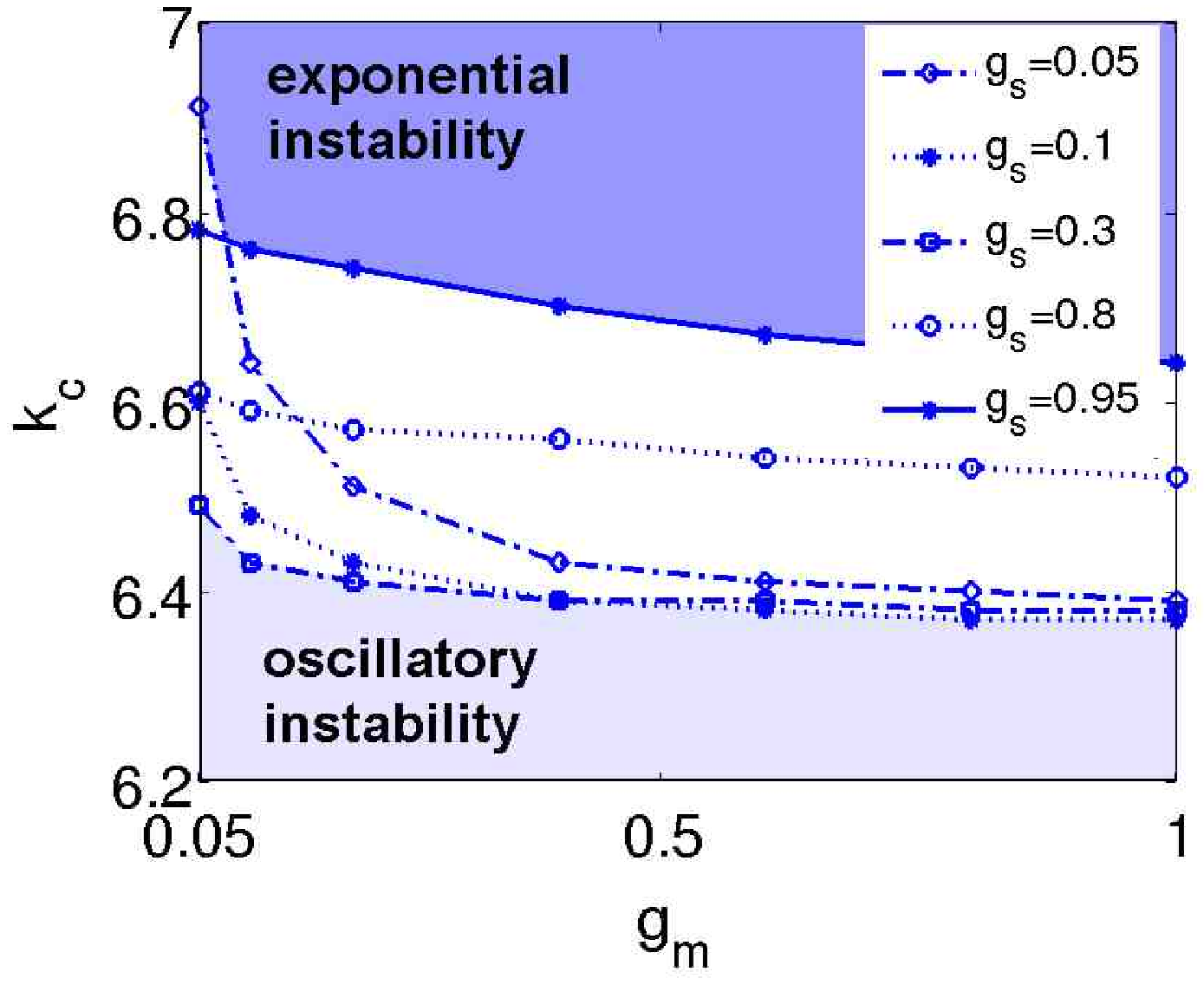}
\caption{[Color online] Dependence of the critical wavenumber $k_c$ on lattice parameter values for aligned nonlinear lattices. The top panel shows $k_c$ versus $g_s$ for 
multiple fixed values of $g_m$, and the bottom panel shows $k_c$ versus $g_m$ 
for multiple fixed values of $g_s$.  The panels labeled  ``exponential instability" and  ``oscillatory instability" indicate regimes that exhibit those instabilities for the parameter values examined.  
} \label{kc_align}
\end{figure}

\begin{figure} [h!t]
\centering 
\includegraphics[width=0.5\textwidth]{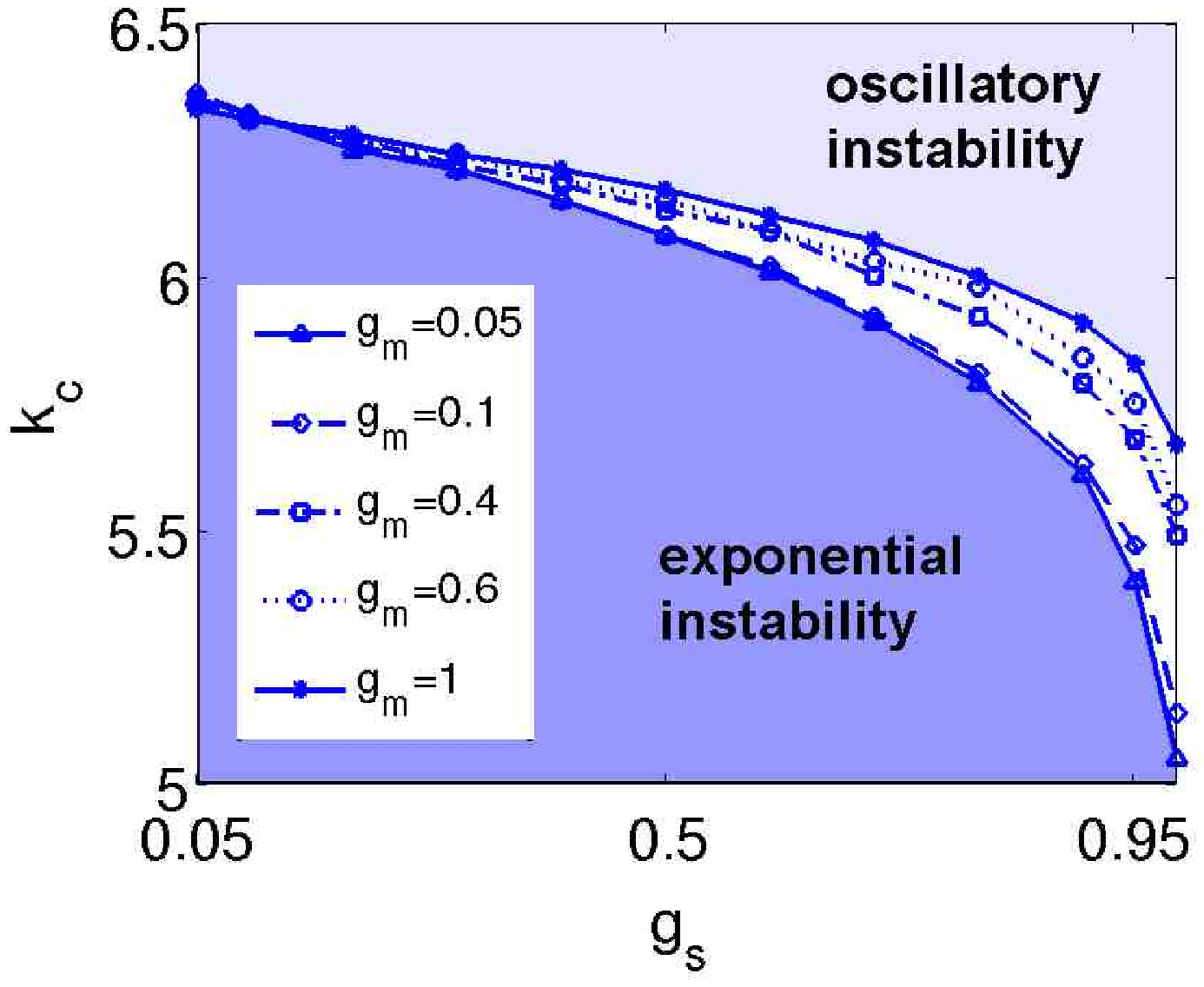}\\
\includegraphics[width=0.5\textwidth]{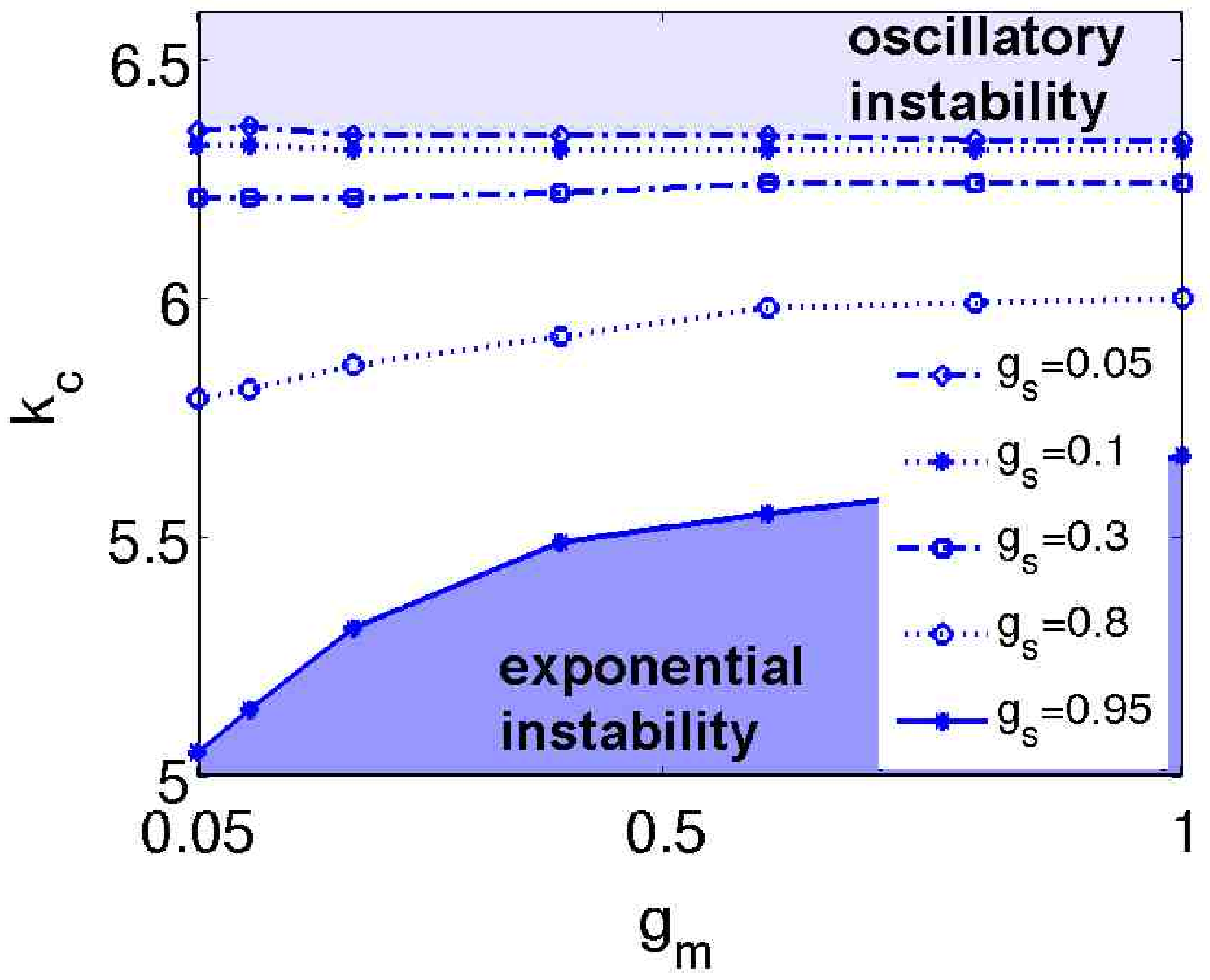}
\caption{[Color online] Dependence of the critical wavenumber $k_c$ on lattice parameter values for anti-aligned nonlinear lattices. The top panel shows $k_c$ versus $g_s$ for multiple fixed values of $g_m$, and the bottom panel shows $k_c$ versus $g_m$ for multiple fixed values of $g_s$. 
}  \label{kc_anti}
\end{figure}

We sweep over the parameter values $g_m \in [0,1]$, $g_s \in[0,0.95]$, 
and $k \in [0,15]$ and thereby confirm, for given $g_s$ and $g_m$, 
that there is a critical wavenumber $k_c$ for both aligned and anti-aligned nonlinear lattices.  For aligned lattices, the anomalous mode increases as a function of $g_m$ away from $g_m = 0$ when $k > k_c$, and it decreases as a function of $g_m$ when $k < k_c$.  For anti-aligned lattices, however, the anomalous mode decreases as a function of $g_m$ away from $g_m = 0$ when $k > k_c$, and it increases as a function of $g_m$ when $k < k_c$. We also trace the critical value $k_c$ between the small-$k$ regime and the large-$k$ regime.  We plot the dependence of $k_c$ on $g_s$ for fixed $g_m$ (top panels) and on $g_m$ for fixed $g_s$ (bottom panels) in Fig.~\ref{kc_align} for the aligned lattice and in Fig.~\ref{kc_anti} for
the anti-aligned lattice.  As one can see from the figures, the transition occurs in a narrow band near $k=6$ for all values of $g_s$ and $g_m$.  Recall
that the healing length (which gives the length scale of the dark solitary wave) is
$\xi=(8\pi N_0 a)^{-1/2}$, where $N_0$ is the maximum
dimensional density of the 3D BEC.  Specifically, $N_0 = n_0 (B \hbar \omega_z / |\tilde{q}|)$, where $\tilde{q} = 4\pi\hbar^2a/m = 2\pi a_\perp^2 q$ and $n_0$ denotes the maximum density of the nondimensional solution. The wavelength of the lattice is $\pi a_z \sqrt{2/B} / k$, so the ratio of the lattice wavelength to the healing length is 
$r_0 \equiv (\pi a_z / k) (16 \pi N_0 a/B)^{1/2}$.  For example, with $k = 6$, the parameter value $n_0 \approx 4$ yields $r_0 \approx 2$.  As we discuss below, this corresponds to the critical region of parameter space, when the width of the dark solitary wave is approximately two lattice wavelengths.  
Hence, when $k$ is small, the variation of the scattering length occurs on a much larger 
scale than that of the solitary wave; when $k$ is
 large, however, the variation occurs at a scale that is smaller than that 
of the solitary wave.  Accordingly, a competition of length scales 
between the scale of the solitary wave and the scale of $g(z)$ accounts for the existence of the two distinct regimes.




For both aligned and anti-aligned lattices, the critical wavenumber $k_c$ exhibits a stronger dependence on $g_s$ than it does on $g_m$, which is consistent with the fact that the depth of the
wells depends increasingly sensitively on $g_s$ as $g_s \rightarrow 1$.  In an aligned lattice with fixed $g_s$, the critical wavenumber $k_c$ decreases with $g_m$, and this decrease becomes more dramatic as $g_s$ becomes smaller.  By contrast, when $g_m$ is fixed, $k_c$ first decreases and then increases with $g_s$; this decrease becomes more dramatic as $g_m$ becomes smaller.  
This variation can be understood intuitively on the basis of
length-scale competition.  Suppose that 
$r_0 \approx 2$ for $k = k_c$.  An increase of $g_m$ then leads to a (maximal) density decrease and hence to a decrease of $k_c$ (because $k_c \propto n_0^{1/2}$ for $r_0 \approx 2$), and the opposite trend emerges from the 
increase of $g_s$.  In an anti-aligned lattice, these trends are reversed (as discussed 
above): with fixed $g_m$, the critical wavenumber $k_c$ decreases with $g_s$, and this decrease becomes more dramatic as $g_m$ becomes smaller.  When $g_s$ is fixed, the BEC dynamics depends on whether $g_s$ is small (i.e., near $0$) or large (i.e., near $1$).  When $g_s$ is small, $k_c$ stays almost constant when $g_m$ changes.  When $g_s$ is large, however, $k_c$ increases with $g_m$.

\section{Dynamical Response}\label{response}

In Section \ref{dark}, we examined the stability of dark solitary waves in BECs with a spatially periodic but time-independent nonlinearity coefficient $g(z)$.  We now consider BECs in which $g$ is initially constant but then a Feshbach resonance is subsequently turned on to implement a spatial dependence.  This allows us to consider experimental situations in which the spatial dependence is turned on slowly (i.e., adiabatically) as well as ones in which it is turned on abruptly (i.e., nonadiabatically).

We study the time-dependent GP equation~(\ref{eq:basicGP}) with $g = g(z,t)$ and an initial wavefunction given by a stationary solution to a GP equation with constant nonlinearity coefficient $g=g_0$. The time-dependent and space-dependent nonlinearity coefficient is given by
\begin{align}
     g(z,t) &= g_0 + \Delta g(z) T(t)\,, \notag \\
     T(t) &= \frac{1}{2} \left[1+\tanh\left(\frac{t-t_0}{\tau} \right)\right]\,, \label{modelg}
\end{align}
where $t_0$ denotes the time at which the function $T(t)$ reaches the
value $1/2$. The function $T(t)$ controlling the transition satisfies $T(t) \rightarrow 0$ as $t \rightarrow 0$ and $T(t) \rightarrow +1$ as $t \rightarrow \infty$.  For each transition time scale $\tau$, we choose $t_0 > 20 \tau$, which ensures that $|T(t)|$ is smaller than machine precision (and hence approximately $0$)
 at $t = 0$ for our computations.  This allows us to model the effect of the nonlinearity coefficient change from an initial value of $g_0$ to a final value of $g_0 + \Delta g(z)$, where $\Delta g(z)$ [see Eq.~(\ref{modeldeltag})] represents the spatial dependence introduced by the Feshbach resonance. A small value of the parameter $\tau$ corresponds to a situation in which the variation of the scattering length is abrupt, whereas a large value of $\tau$ corresponds to a situation in which the Feshbach resonance is turned on gradually (i.e., adiabatically). 



\begin{figure*} [ht]
\begin{minipage}{1\textwidth}
\includegraphics[width=0.3\textwidth]{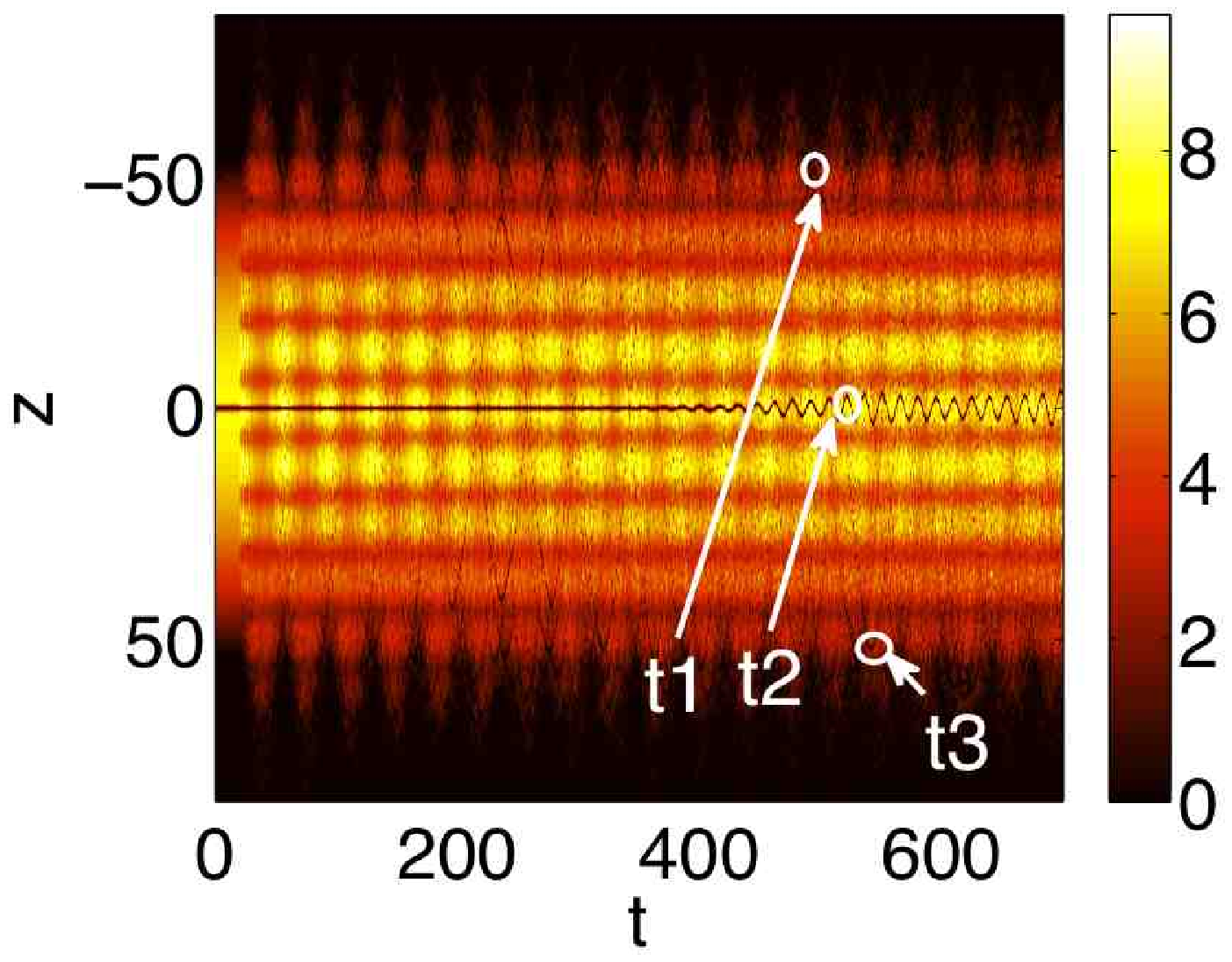}
\includegraphics[width=0.3\textwidth]{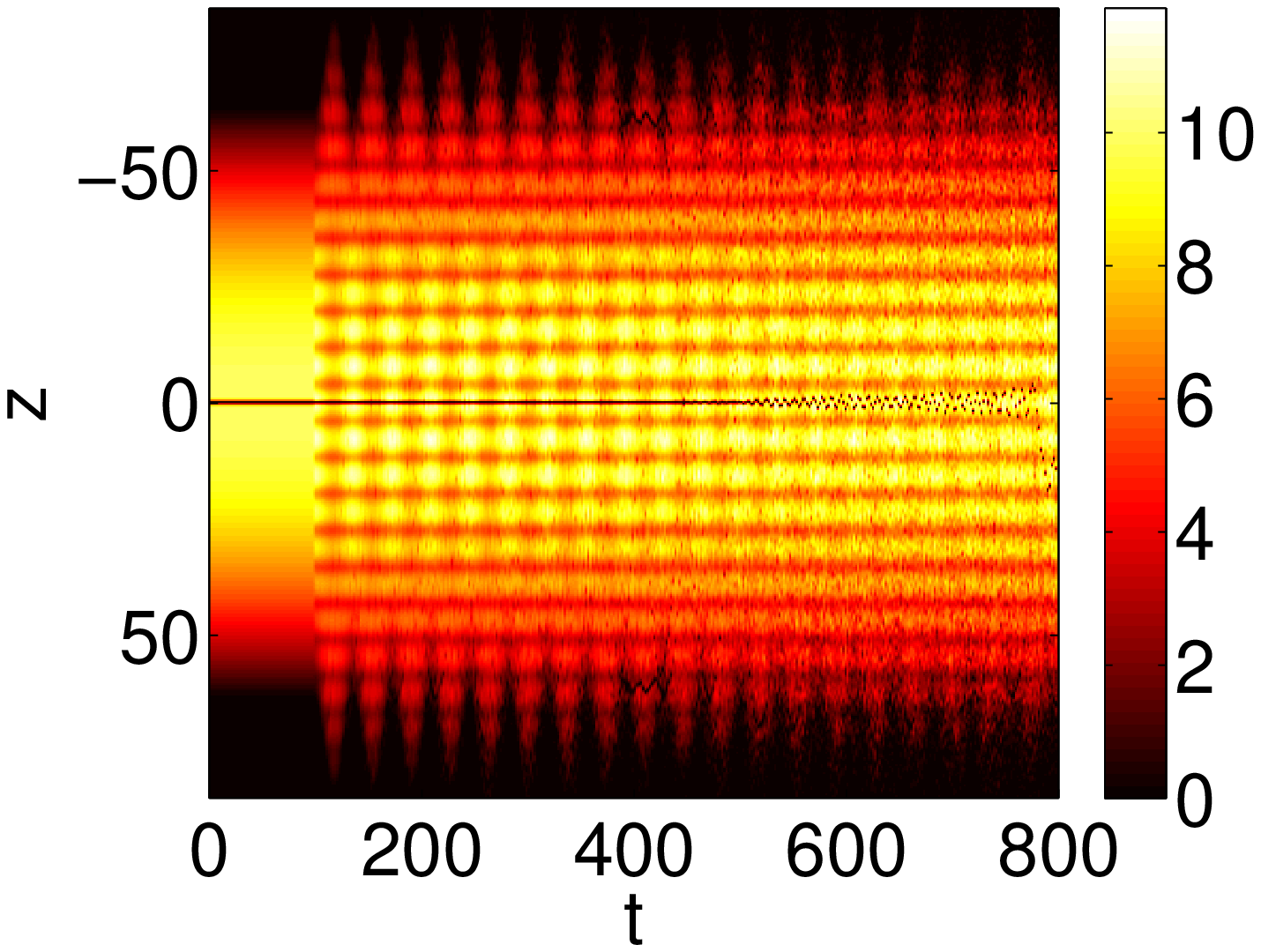}
\includegraphics[width=0.3\textwidth]{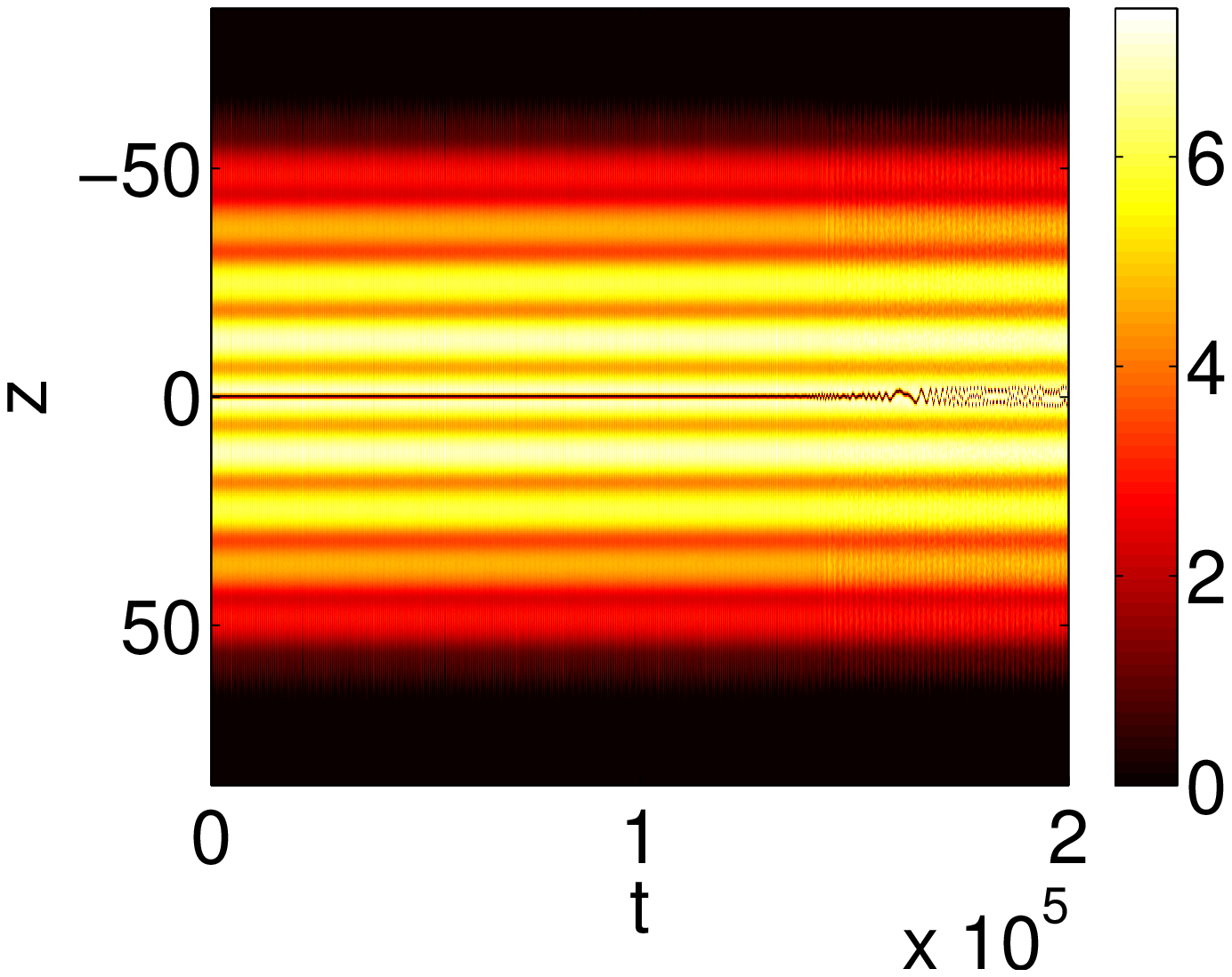}\\
\includegraphics[width=0.3\textwidth]{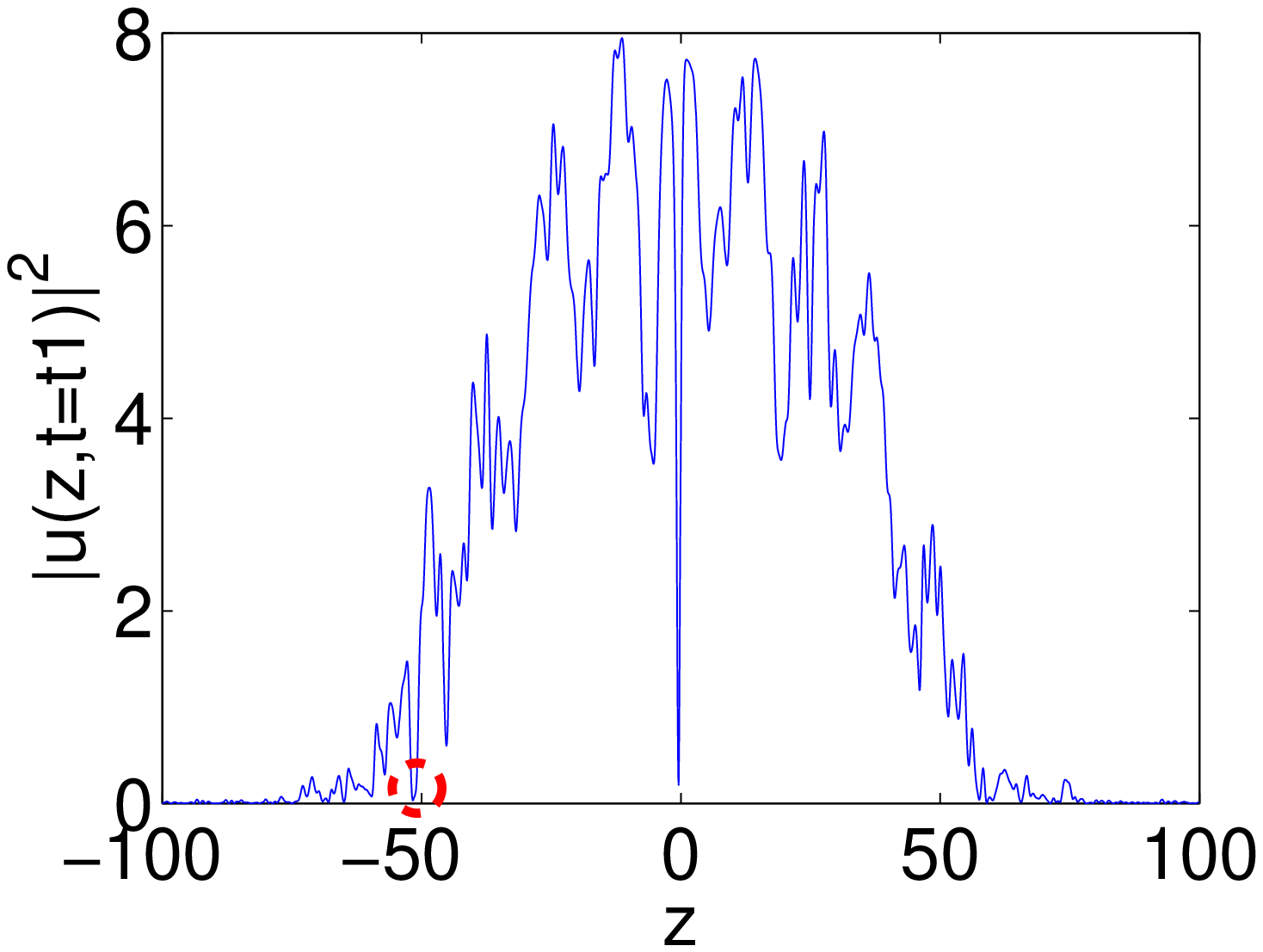}
\includegraphics[width=0.3\textwidth]{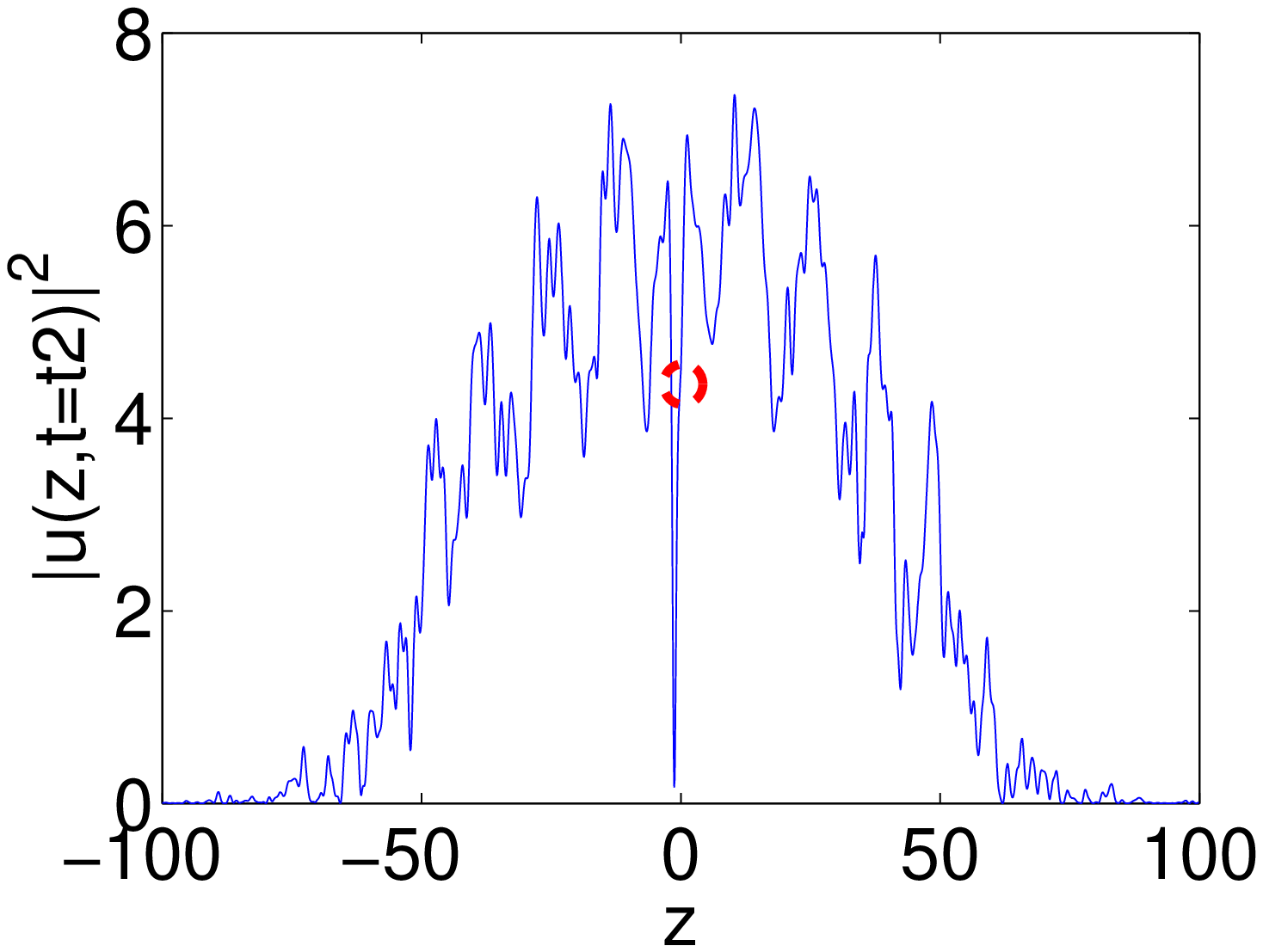}
\includegraphics[width=0.3\textwidth]{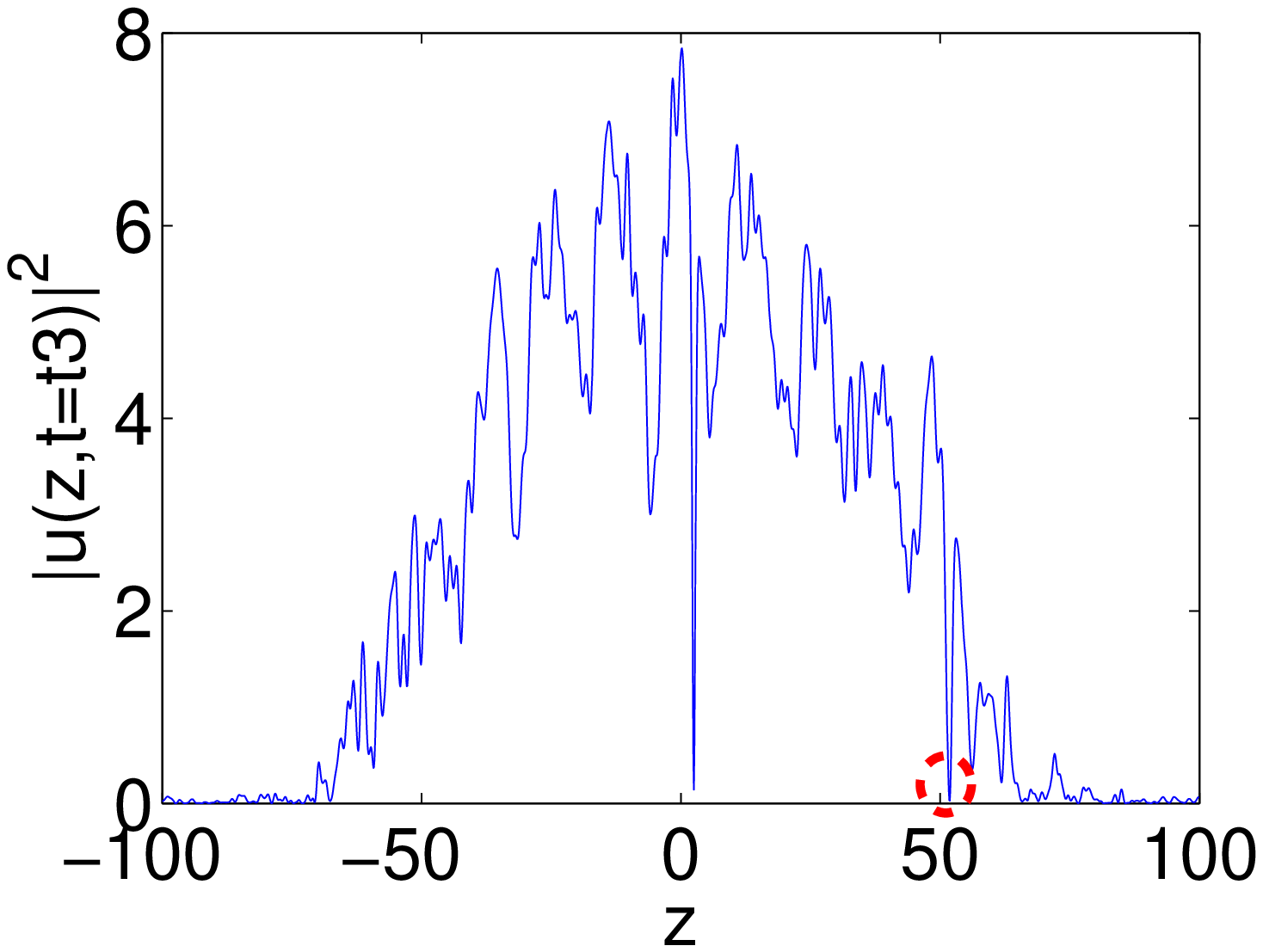}
\end{minipage}
\caption{[Color online] Dynamical response of dark solitary waves in BECs from introducing spatial heterogeneity into the nonlinearity coefficient by turning on a Feshbach resonance. From left to right, the columns correspond to simulations with transition time scales of $\tau=0.1$ (nonadiabatic), $\tau=1$ (medium), and $\tau=10$ (adiabatic). The parameter values describing the shape of the nonlinear lattice are $g_m=0.6$, $g_s=0.5$, $k=0.5$, and $\mu \approx 7.16$. In the left panel of the first row, one can observe the emission of travelling gray solitary waves. We track one such wave: it reaches one of its left turning points (i.e., a point at which the wave's dip is at a value of $z < 0$ such that $|z|$ is a local maximum) at $t_1 \approx 495$, it then crosses $z = 0$ at $t_2 \approx 518$, and it reaches its next right turning point at $t_3 \approx 541$. In the second row, we plot the corresponding spatial profiles of the solitary-wave solution of  the GP equation at times $t_1$ (left panel), $t_2$ (center), and $t_3$ (right).  
In the bottom panels, we circle the gray solitary that we track using a red dashed curve. The local minimum it represents is visible near $z=-50$ in the left column and near $z=50$ in the right column.  (It is contained within the dark solitary wave in the center column.)
 }
\label{dynamics}
\end{figure*}

In Fig.~\ref{dynamics}, we consider the transition time scales $\tau=0.1$, $\tau = 1$, and $\tau = 10$ to illustrate fast, medium, and slow implementation of the spatial inhomogeneity in an aligned nonlinear lattice.  Dark solitary waves appear to persist generically in our dynamical simulations. Nevertheless, unless the onset of the lattice is sufficiently adiabatic (i.e., for $\tau=10$), additional excitations also arise. In particular, when the nonlinear lattice is turned on sufficiently fast (see the panels corresponding to $\tau=0.1$ and $\tau=1$), we observe the emission of localized yet mobile gray solitary waves.  Furthermore, for all  three values of $\tau$, we observe oscillations of the dark solitary wave when we simulate long enough. For $\tau=1$ and $\tau = 10$, the oscillations start at $t \approx 580$ and $t \approx 1.4\times 10^5$, respectively.  This occurs at approximately $t = t_0 + 20\tau$, which is when the nonlinear lattice has settled, within machine precision, to its final form. 
Dark solitary waves are stable if the nonlinearity coefficient is stationary and in its final state of $g = g_0+\Delta g(z)$.



\begin{figure} [h!t]
\centering 
\includegraphics[width=0.5\textwidth]{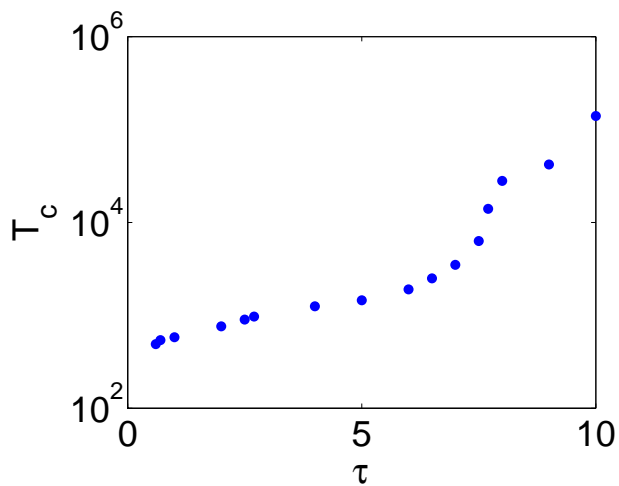}
\caption{[Color online] Onset time $T_c$ for oscillations of a dark solitary wave versus the lattice transition time scale $\tau_c$. 
} \label{tauvsT}
\end{figure}

The eventual mobility of the central solitary wave can be explained by its interaction with sound waves that are emitted during wave propagation \cite{sound}. The strength of the sound waves depends on the value of $\tau$. As $\tau$ becomes larger, the system becomes more adiabatic, so the sound excitations become weaker and it takes longer for the above mechanism to destabilize the solitary wave. Given our collisionally inhomogeneous setup, eventual destabilization is inevitable. We plot the onset time of oscillations $ T_c=T_c(\tau)$ in Fig.~\ref{tauvsT}. By monitoring whether the oscillation has started before a given observation time $T_c$, we can estimate a corresponding threshold time scale $\tau_c$ between nonadiabatic transitions and adiabatic transitions. If the transition time is slower than $\tau_c$, then no substantial oscillations occur before time $t = T_c$. For $T_c=500$, $T_c = 1000$, and $T_c = 5000$, we obtain respective threshold time scales of $\tau_c \approx 0.6$, $\tau_c \approx 2.8$, and $\tau_c \approx 7.4$. We test a series values of $\tau$ and plot the associated $T_c$ values in Fig.~\ref{tauvsT}. 


\section{Conclusions} \label{conclusion}

In this paper, we have studied the structure, stability, and dynamics of dark solitary waves in collisionally inhomogeneous Bose-Einstein condensates.  We considered spatially periodic scattering lengths (i.e., nonlinear lattices) in a functional form suggested by recent experiments.  Importantly, this family of nonlinear lattices can be tuned from a small-amplitude, approximately sinusoidal structure to a periodic sequence of densely-spaced spikes.  We demonstrated several interesting phenomena---including, for example, that dark solitary waves in aligned lattices and anti-aligned lattices exhibit different instability properties. These instability properties
depend significantly on the modulation wavenumber of the nonlinear lattice, and the BEC dynamics exhibit considerable variation when the lattice wavelength is comparable to the size
of the dark solitary waves.

In the case of aligned lattices, we also examined the dynamical response of solitary waves to adiabatic and nonadiabatic implementation of collisional inhomogeneities via a Feshbach 
resonance.  When a Feshbach resonance is turned on nonadiabatically, nonlinear excitations can be emitted.  Sound waves are emitted when a Feshbach resonance is turned on (regardless of the transition speed), and these excitations eventually destablize the solitary wave.  If the resonance is turned on sufficiently adiabatically, such that the sound waves are weak, then the BEC settles to an excited state (bearing a dark solitary wave) of the stationary nonlinear lattice initially before eventually starting to oscillate after a very long time.


It would be interesting to generalize our considerations to 
higher-dimensional lattices and to examine the effects that such lattices
have on higher-dimensional excitations, such as vortices in quasi-2D
BECs and vortex rings in 3D condensates.
Although some initial efforts have been made in that direction~\cite{jpb},
a systematic theory has yet to be developed, and thorough numerical investigations
would also be beneficial. 


\section*{Acknowledgements}

CW thanks St. Catherine's College at Oxford for funding from a College Scholarship in the sciences. KJHL acknowledges EPSRC and ONR for financial support. PGK gratefully acknowledges support from the US National Science Foundation through grant DMS-0806762 and from the Alexander von Humboldt Foundation. We thank Ian Spielman for many helpful discussions on this topic and for bringing Ref.~\cite{takahashi} to our attention.  We also thank Alexandru Nicolin and an anonymous referee for helpful comments.



\begin{thebibliography}{99}

\bibitem{book1} C.~J. Pethick and H. Smith,
{\it Bose-Einstein Condensation in Dilute Gases}, Second Edition,
Cambridge University Press (Cambridge, 2008).

\bibitem{book2} L.~P. Pitaevskii and S. Stringari,
{\it Bose-Einstein Condensation}, Oxford University Press (Oxford, 2003).

\bibitem{expb1} K.~E.\ Strecker, G.~B.\ Partridge, A.~G.\ Truscott, and R.~G.\ Hulet,
Nature \textbf{417}, 150 (2002).

\bibitem{expb2} L.\ Khaykovich, F.\ Schreck, G.\ Ferrari, T.\ Bourdel, J.\ Cubizolles, L.~D.\ Carr,
Y.\ Castin, and C.\ Salomon, Science \textbf{296}, 1290 (2002).

\bibitem{expb3}
S.~L. Cornish, S.~T. Thompson, and C.~E. Wieman,
Phys.\ Rev.\ Lett.\ {\bf 96}, 170401 (2006).

\bibitem{dark1} S.\ Burger, K.\ Bongs, S.\ Dettmer, W.\ Ertmer, K.\ Sengstock, A.\ Sanpera,
G.~V.\ Shlyapnikov, and M.\ Lewenstein,
Phys.\ Rev.\ Lett.\ \textbf{83}, 5198 (1999).

\bibitem{dark2}
J.\ Denschlag, J.~E.\ Simsarian, D.~L. \ Feder, C.~W.\ Clark, L. A.\ Collins, J.\ Cubizolles, L.\ Deng,
E. W.\ Hagley, K.\ Helmerson, W. P.\ Reinhardt, S. L.\ Rolston, B. I.\ Schneider, and W. D.\ Phillips,
Science \textbf{287}, 97 (2000).

\bibitem{dark3}
B.~P.\ Anderson, P. C.\ Haljan, C. A.\ Regal, D. L.\ Feder, L. A.\ Collins, C. W.\ Clark, and E. A.\ Cornell, Phys.\ Rev.\ Lett.\ \textbf{86}, 2926 (2001).

\bibitem{dark4}
Z.\ Dutton, M.\ Budde, Ch.\ Slowe, and L. V.\ Hau,
Science \textbf{293}, 663 (2001).

\bibitem{dark5} P. Engels and C. Atherton, Phys. Rev. Lett. {\bf 99}, 
160405 (2007). 


\bibitem{dark6} A. Weller, J. P. Ronzheimer, C. Gross, J. Esteve, M. K. Oberthaler, D. J. Frantzeskakis, G. Theocharis, and P. G. Kevrekidis
Phys. Rev. Lett. {\bf 101}, 130401 (2008); G. Theocharis, A. Weller, J. P. Ronzheimer, C. Gross, M. K. Oberthaler, P. G.  Kevrekidis, and D. J. Frantzeskakis
Phys. Rev. A {\bf 81}, 063604 (2010).

\bibitem{dark7} I. Shomroni, E. Lahoud, S. Levy, and J. Steinhauer, 
Nature Phys. {\bf 5}, 193 (2009). 

\bibitem{dark8} 
M. Baumert, E.-M. Richter, J. Kronj{\"a}ger, K. Bongs, and K. Sengstock, Nature Phys. 4, 496 (2008);  
S. Stellmer, C. Becker, P. Soltan-Panahi, E.-M. Richter, S. D{\"o}rscher, M. Baumert, J. Kronj{\"a}ger, K. Bongs, and K. Sengstock, Phys. Rev. Lett. {\bf 101}, 120406 (2008). 

\bibitem{gap} B.\ Eiermann, Th.\ Anker, M.\ Albiez, M.\ Taglieber, P.\ Treutlein, K.-P.\ Marzlin, and M. K.\ Oberthaler, Phys.\ Rev.\ Lett.\ \textbf{92}, 230401 (2004).


\bibitem{db1} C. Becker, S. Stellmer, P. Soltan-Panahi, S. D{\"o}rscher, 
M. Baumert, E.-M. Richter, J. Kronj{\"a}ger, K. Bongs, and K. Sengstock, Nature Phys. {\bf 4}, 496 (2008).

\bibitem{db2} S. Middelkamp, J. J. Chang, C. Hamner, R. Carretero-Gonz{\'a}lez, P. G. Kevrekidis, V. Achilleos, D. J. Frantzeskakis, P. Schmelcher, and P. Engels, Phys. Lett. A {\bf 375}, 642 (2011). 

\bibitem{db3} C. Hamner, J. J. Chang, P. Engels, and M. A. Hoefer
Phys. Rev. Lett. 106, 065302 (2011). 

\bibitem{db4} D. Yan, J. J. Chang, C. Hamner, P. G. Kevrekidis, P. Engels, V. Achilleos, D. J. Frantzeskakis, R. Carretero-Gonz{\'a}lez, and P. Schmelcher
Phys. Rev. A {\bf 84}, 053630 (2011). 

\bibitem{hoefer} M. A. Hoefer, J. J. Chang, C. Hamner, and P. Engels
Phys. Rev. A {\bf 84}, 041605 (2011). 



\bibitem{ourbook} P. G. Kevrekidis, D. J. Frantzeskakis, and R. Carretero-Gonz\'alez (eds.),
{\it Emergent Nonlinear Phenomena in Bose-Einstein Condensates. Theory and Experiment}, Springer-Verlag (Berlin, 2008).

\bibitem{review} R.\ Carretero-Gonz\'alez, D.~J.\ Frantzeskakis, and P.~G.\ Kevrekidis,
Nonlinearity {\bf 21}, 139 (2008).

\bibitem{engels07} P. Engels, C. Atherton, and M.~A. Hoefer, Phys. Rev. Lett. {\bf 98}, 095301 (2007). 

\bibitem{alex1} A.~I. Nicolin, Phys. Rev. E {\bf 84}, 056202 (2011).

\bibitem{alex2} A. Bala\v{z} and A.~Nicolin, Phys. Rev. A {\bf 85}, 023613 (2012).

\bibitem{pgk} P.~G. Kevrekidis and D.~J. Frantzeskakis,
Mod. Phys. Lett. B {\bf 18}, 173 (2004).

\bibitem{konotop} V.~A. Brazhnyi and V.~V. Konotop,
Mod. Phys. Lett. B {\bf 18}, 627 (2004).

\bibitem{blochol}
I. Bloch, J.\ Phys.\ B: At.\ Mol.\ Opt.\ Phys.\ {\bf 38}, S629 (2005); I. Bloch, Nature Phys. {\bf 1}, 23 (2005).

\bibitem{morsch}
O. Morsch and M. Oberthaler, Rev. Mod. Phys. {\bf 78}, 179 (2006).

\bibitem{Folman}  R.\ Folman, P.\ Kr\"uger, J.\ Schmiedmayer, J.\ Denschlag, and C.\ Henkel, Adv.\ Atom.\ Mol.\ Opt.\ Phys.\ {\bf 48}, 263 (2002);
J. Reichel, Appl. Phys. B {\bf{75}}, 469 (2002);
J. Fortagh and C. Zimmermann, Science  {\bf{307}} 860 (2005);
J. Fortagh and C. Zimmermann, Rev. Mod. Phys. {\bf{79}}, 235 (2007).

\bibitem{Grimm}
R. Grimm, M. Weidem\"uller, and Y. B. Ovchinnikov, Adv. At. Mol. Opt. Phys. {\bf{42}}, 95 (2000).

\bibitem{Lesanovsky}
S. Hofferberth, I. Lesanovsky, B. Fischer, J. Verdu, and J. Schmiedmayer, Nature Phys. {\bf{2}} 710 (2006);
I. Lesanovsky, S. Hofferberth, J. Schmiedmayer, and P. Schmelcher, Phys. Rev. A {\bf{74}} 33619 (2006);
I. Lesanovsky, T. Schumm, S. Hofferberth, L. M. Andersson, P. Kr\"uger and J. Schmiedmayer, Phys. Rev. A {\bf{73}} 33619 (2006).

\bibitem{Koehler}
T. K\"ohler, K. Goral, and P. S. Julienne, Rev. Mod. Phys. {\bf{78}}, 1311 (2006).
 
\bibitem{feshbachNa} S. Inouye, M. R. Andrews, J. Stenger, H. J. Miesner, D. M. Stamper-Kurn, and W. Ketterle, Nature {\bf 392}, 151 (1998); J. Stenger, S. Inouye, M. R. Andrews, H.-J. Miesner,
D. M. Stamper-Kurn, and W. Ketterle, Phys. Rev. Lett. {\bf 82}, 2422 (1999);
J. L. Roberts, N. R. Claussen, J. P. Burke Jr., C. H. Greene, E. A. Cornell, and
C.~E. Wieman, Phys. Rev. Lett. {\bf 81}, 5109 (1998); S.~L. Cornish,
N. R. Claussen, J. L. Roberts, E. A. Cornell, and C. E. Wieman,
Phys. Rev. Lett. {\bf 85}, 1795 (2000).
 
\bibitem{ofr} F. K. Fatemi, K. M. Jones, and P. D. Lett, Phys. Rev. Lett. {\bf 85}, 4462 (2000);
M. Theis, G. Thalhammer, K. Winkler, M. Hellwig, G. Ruff, R. Grimm, and J. H. Denschlag,
Phys. Rev. Lett. {\bf{93}}, 123001 (2004).



\bibitem{molecule} J. Herbig,
T. Kraemer, M. Mark, T. Weber, C. Chin, H. C. Nagerl, and R. Grimm,  Science \textbf{301}, 1510 (2003); C.~A. Regal, C. Ticknor, J. L. Bohn, and D. S. Jin, Nature \textbf{424}, 47 (2003).

\bibitem{becbcs} M. Bartenstein,
A. Altmeyer, S. Riedl, S. Jochim, C. Chin, J. H. Denschlag, and R. Grimm,
Phys. Rev. Lett. \textbf{92}, 203201 (2004); T. Bourdel, L. Khaykovich, J. Cubizolles, J. Zhang, F. Chevy, M. Teichmann, L. Tarruell, S. J .J. M. F. Kokkelmans, and C. Salomon, Phys. Rev. Lett. {\bf 93}, 050401 (2004).

\bibitem{FRM1} F.~Kh. Abdullaev, J. G. Caputo, R. A. Kraenkel, and B. A. Malomed, 
Phys. Rev. A {\bf 67}, 013605 (2003); H. Saito and M. Ueda, Phys. Rev. Lett. {\bf 90}, 040403 (2003);
G.~D. Montesinos, V. M. P\'erez-Garc\'{\i}a, and P. J. Torres, Physica D {\bf 191} 193 (2004).

\bibitem{FRM2} P.~G. Kevrekidis, G. Theocharis, D. J. Frantzeskakis, and B. A. Malomed,
Phys. Rev. Lett. {\bf 90}, 230401 (2003);
Z.~X. Liang,  Z. D. Zhang, and W. M. Liu, Phys. Rev. Lett. {\bf 94}, 050402 (2005);
M. Matuszewski, E. Infeld, B. A. Malomed, and M. Trippenbach, Phys. Rev. Lett. {\bf 95}, 050403 (2005).

\bibitem{martin} M. Centurion, M. A. Porter, P. G. Kevrekidis, and D. Psaltis,
Phys. Rev. Lett. {\bf 97}, 033903 (2006); M. Centurion, M. A. Porter,
Y. Pu, P. G. Kevrekidis, D. J. Frantzeskakis, and D. Psaltis,
Phys. Rev. Lett. {\bf 97}, 234101 (2006); M. Centurion, M. A. Porter,
Y. Pu, P. G. Kevrekidis, D. J. Frantzeskakis, and D. Psaltis,
Phys. Rev. A {\bf 75}, 063804 (2007);
S. Beheshti, K. J. H. Law, P. G. Kevrekidis, and M. A. 
Porter, Phys. Rev. A {\bf 78}, 025805 (2008).

\bibitem{dep1} V. Zharnitsky and D. E. Pelinovsky, Chaos \textbf{15}, 037105 (2005).

\bibitem{dep2} D.~E. Pelinovsky, P. G. Kevrekidis, and D. J. Frantzeskakis, Phys. Rev. Lett. {\bf 91}, 240201 (2003);
D.~E. Pelinovsky, P. G. Kevrekidis, D. J. Frantzeskakis, and V. Zharnitsky, Phys. Rev. E {\bf 70}, 047604 (2004).

\bibitem{dep3}  P.~G. Kevrekidis, D.~E. Pelinovsky, and A. Stefanov,
J. Phys. A {\bf 39}, 479 (2006).

\bibitem{our1} G. Theocharis, P. Schmelcher, P. G. Kevrekidis, and D.J. Frantzeskakis,
Phys. Rev. A {\bf 72}, 033614 (2005).

\bibitem{fka} F. Kh. Abdullaev and M. Salerno, J. Phys. B {\bf 36}, 2851 (2003).

\bibitem{our2} G. Theocharis, P. Schmelcher, P. G. Kevrekidis, and D. J. Frantzeskakis,
Phys. Rev. A {\bf 74}, 053614 (2006).

\bibitem{fka2} J. Garnier and F. Kh. Abdullaev, Phys. Rev. A {\bf 74}, 013604 (2006).


\bibitem{LocDeloc} Yu.~V. Bludov, V.~A. Brazhnyi, and V.~V. Konotop, Phys. Rev. A {\bf 76}, 023603 (2007).

\bibitem{yiota} P. Niarchou, G. Theocharis, P. G. Kevrekidis, P. Schmelcher, and D. J. Frantzeskakis, Phys. Rev. A {\bf 76}, 023615 (2007).

\bibitem{vpg14} F.~Kh. Abdullaev and J. Garnier, Phys. Rev. A {\bf 72}, 061605(R) (2005).

\bibitem{vpg16} H. Sakaguchi and B.~A. Malomed, Phys. Rev. E {\bf 72}, 046610 (2005); M.~A. Porter, P. G. Kevrekidis, B. A. Malomed, and D. J. Frantzeskakis, Physica D {\bf 229}, 104 (2007);
F. Kh. Abdullaev, A. Abdumalikov and R. Galimzyanov, Phys. Lett. A {\bf 367}, 149 (2007).

\bibitem{BludKon} Y.~V. Bludov and V.~V. Konotop, Phys. Rev. A {\bf 74},
043616 (2006).

\bibitem{rodrig08} A.~S. Rodrigues, P.~G. Kevrekidis, M.~A. Porter, D.~J. Frantzeskakis, P. Schmelcher, and A.~R. Bishop, Phys. Rev. A {\bf 78}, 013611 (2008).

\bibitem{vpg12} M.~I. Rodas-Verde, H. Michinel,
and V.~M. P{\'e}rez-Garc{\'i}a, Phys. Rev. Lett. {\bf 95}, 153903 (2005);
A. V. Carpentier, H. Michinel, M. I. Rodas-Verde,
and V. M. P\'erez-Garc\'{\i}a, Phys. Rev. A {\bf 74}, 013619 (2006).

\bibitem{vpg17} M.~T. Primatarowa, K.~T. Stoychev, and R.~S. Kamburova, Phys. Rev. E {\bf 72}, 036608 (2005).

\bibitem{vpg_new} V.~M. P{\'e}rez-Garc{\'i}a, arXiv:nlin/0612028.

\bibitem{key-2} G. Fibich, Y. Sivan, and M.~I. Weinstein, Physica D {\bf 217}, 31 (2006).

\bibitem{key-4} Y. Sivan, G. Fibich, and M.~I. Weinstein, Phys. Rev. Lett. {\bf 97}, 193902 (2006);

\bibitem{vprl} J. Belmonte-Beitia, V. M. P\'erez-Garc\'{\i}a, V. Vekslerchik, and P. J. Torres,
Phys. Rev. Lett. {\bf 98}, 064102 (2007).

\bibitem{kominis} Y. Kominis and K. Hizanidis, Opt. Express {\bf 16}, 12124 (2008).

\bibitem{ckrtj} Z. Rapti, P. G. Kevrekidis, V. V. Konotop, and C. K. R. T. Jones, J. Phys. A: Math. Theor. {\bf 40}, 14151 (2007).

\bibitem{blud_pre} F.~Kh. Abdullaev, Yu. V. Bludov, S. V. Dmitriev, P. G. Kevrekidis, and V. V. Konotop, Phys. Rev. E {\bf 77}, 016604 (2008).



\bibitem{malomed} Y. V. Kartashov, B. A. Malomed, and L. Torner, Rev. Mod. Phys. {\bf 83}, 247 (2011). 

\bibitem{takahashi} R. Yamazaki, S. Taie, S. Sugawa, Y. Takahashi,
Phys. Rev. Lett. {\bf 105}, 050405 (2010).

\bibitem{jpb} S. Middelkamp, P. G. Kevrekidis, D. J. Frantzeskakis,
R. Carretero-Gonz{\'a}lez, P. Schmelcher, 
J. Phys. B {\bf 43}, 155303 (2010).

\bibitem{ds_dimitri} D.~J. Frantzeskakis, J. Phys. A {\bf 43}, 213001 (2010).

\bibitem{sound} N. G. Parker, N. P. Proukakis, C. S. Adams, Phys. Rev. A {\bf 81}, 033606 (2010).


\end{thebibliography}
\end{document}